\documentclass{aa}  
\usepackage{graphicx}
\usepackage{txfonts}
\usepackage[colorlinks=True,allcolors=blue]{hyperref}

\usepackage{mathrsfs}
\usepackage{amsbsy}

\usepackage{pifont}
\usepackage{subcaption}
\begin{document}

   \title{APEX CO observations towards the photodissociation region of RCW~120}

   \subtitle{}

   \author{M. Figueira 
          \inst{1}
          \and
          A. Zavagno\inst{2}
          \and
          L. Bronfman\inst{3}
          \and
          D. Russeil\inst{2}
          \and
          R. Finger\inst{3}
          \and
          F. Schuller\inst{4,5}
          }
   \institute{National Centre for Nuclear Research, ul. Pasteura 7, 02-093, Warszawa, Poland
         \and
             Aix Marseille Univ, CNRS, CNES, LAM, Marseille, France
         \and
             Departamento de Astronomía, Universidad de Chile, Casilla 36-D, Santiago, Chile
         \and
             Max-Planck-Institut für Radioastronomie, Auf dem Hügel 69, 53121, Bonn, Germany
         \and
             Leibniz-Institut f\"{u}r Astrophysik Potsdam (AIP), An der Sternwarte 16, 14482 Potsdam, Germany
             }

   \date{Received **, ****; accepted **, ****}

 
  \abstract
   {The edges of ionized (H{\,\sc{ii}}) regions are important sites for the formation of (high-mass) stars. Indeed, at least 30\% of the galactic high mass star formation is observed there. The radiative and compressive impact of the H{\,\sc{ii}} region could induce the star formation at the border following different mechanisms such as the Collect \& Collapse (C\&C) or the Radiation Driven Implosion (RDI) models and change their properties.}
   {We study the properties of two zones located in the Photo Dissociation Region (PDR) of the Galactic H{~\sc{ii}} region RCW~120 and discussed them as a function of the physical conditions and young star contents found in both clumps.}
   {Using the APEX telescope, we mapped two regions of size 1.5'$\times$1.5' toward the most massive clump of RCW~120 hosting young massive sources and toward a clump showing a protrusion inside the H{\,\sc{ii}} region and hosting more evolved low-mass sources. The $^{12}$CO~($J=3-2$), $^{13}$CO~($J=3-2$) and C$^{18}$O~($J=3-2$) lines observed, together with {\textit{Herschel}} data are used to derive the properties and dynamics of these clumps. We discuss their relation with the hosted star-formation.}
   {Assuming LTE, the increase of velocity dispersion and $T_{ex}$ are found toward the center of the maps, where star-formation is observed with \textit{Herschel}. Furthermore, both regions show supersonic Mach number (7 and 17 in average). No strong evidences have been found concerning the impact of far ultraviolet (FUV) radiation on C$^{18}$O photodissociation at the edges of RCW~120. The fragmentation time needed for the C\&C to be at work is equivalent to the dynamical age of RCW~120 and the properties of region B are in agreement with bright-rimmed clouds.}
   {Despite that the conclusion from this fragmentation model should be taken with caution, it strengthens the fact that, together with evidences of compression, C\&C might be at work at the edges of RCW~120. Additionally, the clump located at the eastern part of the PDR is a good candidate of pre-existing clump where star-formation may be induced by the RDI mechanism.}


   \maketitle
%

\section{Introduction}

High-mass stars ($M\ge 8M_{\odot}$) have a strong impact on the interstellar medium (ISM), galaxies formation and evolution. From the radiation field, ionization pressure, to the explosion as supernovae, they strongly shape their surroundings, due to high energy and momentum budgets of this feedback, and release metals \citep{kru14,gee19}. Therefore, while high-mass stars represent a minor part of the stellar population, the consequences of their feedback are primordial. In the study of star formation, there is one particular structure which directly relates the feedback of high-mass stars to the new generation of stars, and is called an ionized (H{\,\sc{ii}}) region. These objects are created by the ionizing radiation of massive stars \citep{str39} and the further expansion of the H{\,\sc{ii}} region \citep{spi78} due to the temperature difference between the ionized gas ($\sim$8000~K) and the surrounding medium ($\sim$20~K). During this expansion, a layer is formed between the ionization front (IF) and the shock front (SF) that preceeds the IF during the expansion of the region in the surrounding medium. The whole structure is often called an H{\,\sc{ii}} bubble, even though the geometry cannot be easily assessed \citep{bea10, and15}. Using the WISE catalog, \citet{and14} identified 8000 of these H{\,\sc{ii}} regions in the Galactic Plane. When the layer of material surrounding the ionizing stars is dense enough, star formation can be observed in it. This mechanism, where one or several high-mass stars are responsible for star formation is called a triggering mechanism, and is thought to be a plausible explanation for the presence of OB associations \citep{bla64,pre07}. Over the years, two mains models explaining the formation of a new generation of stars due to the expansion of an H{\,\sc{ii}} region emerged. The first one is the Collect \& Collapse (C\&C, \citealt{elm77,whi94b}) process, explaining the creation and fragmentation of the layer of material and the second is the Radiation Driven Implosion (RDI, \citealt{kes03}) model, where the interaction between a pre-existing, stable clump and the H{\,\sc{ii}} region induces the star formation. Simulations of H{~\sc{ii}} regions expansion in a turbulent medium, show the formation of pillars and cometary globules \citep{tre12}; and the expansion in a fractal medium triggers the formation of stars by combining elements from C\&C and RDI \citep{wal15}. Several theoretical \citep{ber89, lef94, mia06} and observational works \citep{urq09,mor09,fuk13} have been performed to study the interaction between a clump and the H{\,\sc{ii}} region through the RDI process. It is associated with an high ionizing flux \citep{bis11}, an elongated tail, ionized boundary layer (IBL) and 8~$\mu$m emission. Some observations have shown that several components can be observed toward Bright Rimmed Clouds (BRCs) due to the internal dynamics caused by the interaction, although the presence of an IBL and/or 8~$\mu$m is often taken as a proof of interaction with the ionizing flux. Statistical studies using \textit{Spitzer}, ATLASGAL and \textit{Herschel} showed that H{\,\sc{ii}} bubbles host 25$-$30\%, at least, of the high-mass Galactic sources \citep{deh10,ken12,ken16,pal17} and dedicated studies also show the same behaviour \citep{tig17,rus19,xu19}. Therefore, this high percentage of young high-mass stars was mainly thought to be the result of different triggering mechanisms. However, simulations of expanding H{\,\sc{ii}} regions mostly showed that stars, whose formation were triggered, are not dominant and that spontaneously formed stars (without any help from stellar feedback) are also found at the edges of H{\,\sc{ii}} regions \citep{dal15}. Additionally, numerical simulations also show that the interaction of the neutral material with the H{\,\sc{ii}} region could have a negative or no impact with respect to star formation \citep{gee15,dal11,dal17} such as lowering the Star Formation Efficiency (SFE) compared to what is expected from observations \citep{gee16,rah19,dal12b} which is therefore not in support of triggering mechanisms.\\

\begin{figure}[h!]
  \centering
  \includegraphics[width=\linewidth]{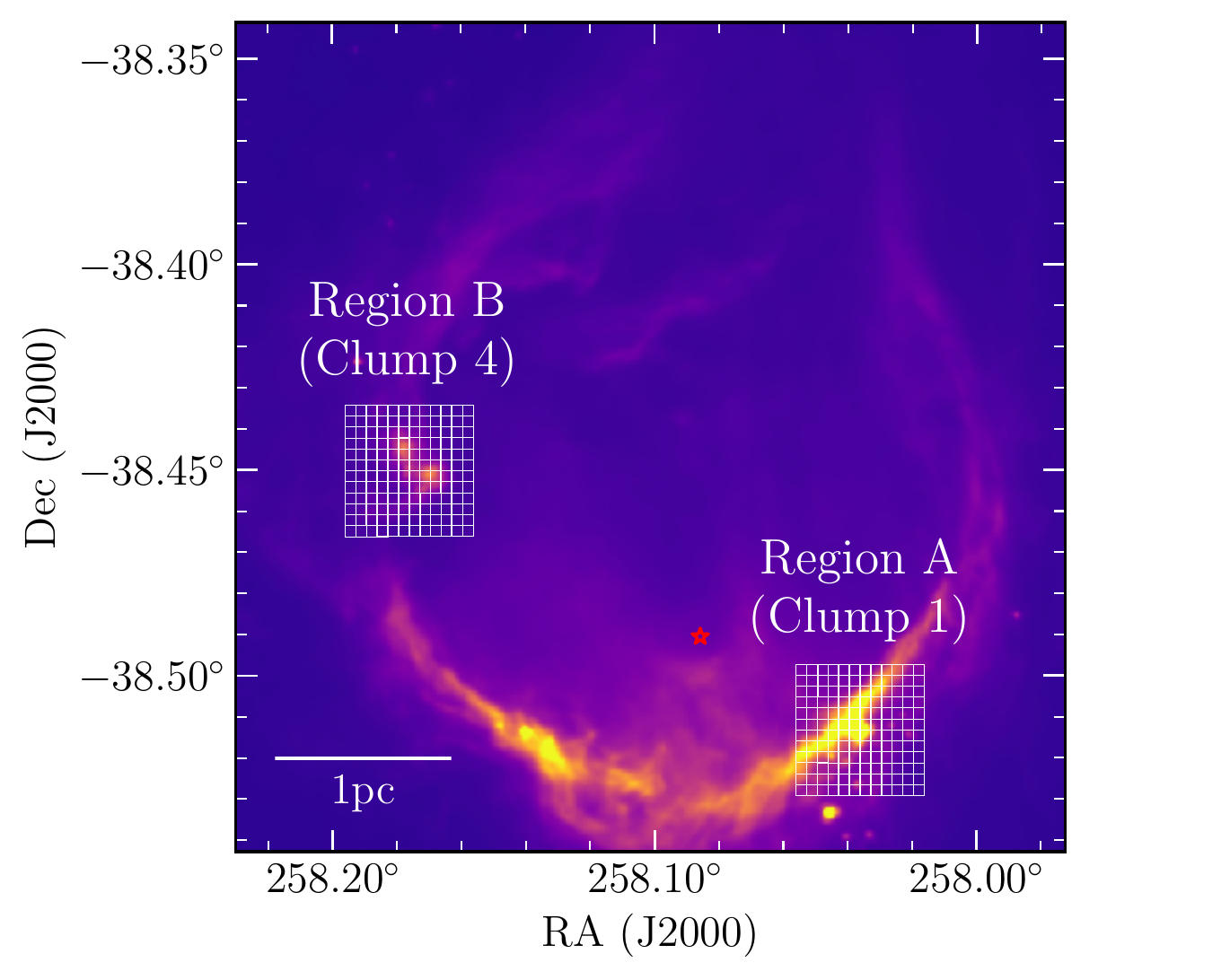}
  \caption{\textit{Herschel} 70~$\mu$m observation of RCW~120. The two grids (in white) represent the APEX observations with 121 ON observations per grid separated by half the beam for a width of 105\arcsec, The ionizing star is represented by the red star symbol.}
  \label{fig:RCW120_APEX_mapping}
\end{figure}

RCW~120 is a well studied Galactic H{\,\sc{ii}} region due to its ovoid shape and its relatively close distance (1.34~kpc, \citealt{rus03,zav07}). Thanks to these advantages, this region has received a lot of attention from observers and simulations, and has been studied in several papers \citep{zav07,deh09,zav10,and12,and15,tre14,kir14,tor15,wal15,mac16,fig17,fig18,mar19,kir19,zav20}. \citet{fig17} showed that two millimeter-wave observed clumps host different kind of sources with respect to their evolutionary stage. These clumps, defined at 1.3~mm in \citet{zav07}, are located in the south west (Clump 1) and middle east (Clump 4),  and respectively covered by region A and region B (see Fig.~\ref{fig:RCW120_APEX_mapping}). In term of YSOs, the clump 1 hosts massive and young sources while the clump 4 hosts low and more evolved sources. Considering the projected distance to the ionizing star, the expansion of the H{\,\sc{ii}} region should have impacted the clump 1 in first place compared to the clump 4. Therefore, since high-mass star formation proceeds faster than the low-mass equivalent \citep{sch15}, we should have therefore found older sources toward the clump 1, which we did not. However, this interpretation has to be taken with caution since age gradients cannot be taken as a strong evidence of triggering \citep{dal13a}. This simple hypothesis does not take into account the mass of the cores which also plays a role in star formation timescales. Additionally, the dust distribution seen at 70~$\mu$m and the PAHs emission at 8~$\mu$m is quite different for the clump 1, a roughly flat layer of material, and the clump 4, a distorted layer of material in V-shape. Hence, it is possible that the interaction between the H{\,\sc{ii}} region and the layer in the south-western and eastern parts is of different nature. Using APEX-SheFI observations of $^{12}$CO, $^{13}$CO and C$^{18}$O in the $J=3-2$ transition, we studied the clumps 1 and 4 of \citet{fig17} covered by region A and region B, respectively (Fig.~\ref{fig:RCW120_APEX_mapping}).\\
In Sect.~2, we present the APEX observations and data reduction which are analyzed in Sect.~3. In Sect.~4, we discuss the results regarding to induced star-formation and we conclude in Sect.~5.

\section{APEX-SHeFI observations and data reduction}

The $^{12}$CO~(3$-$2), $^{13}$CO~(3$-$2) and C$^{18}$O~(3$-$2) molecular line transitions were observed with the Atacama Pathfinder EXperiment\footnote{This publication is based on data acquired with the Atacama Pathfinder Experiment (APEX). APEX is a collaboration between the Max-Planck-Institut fur Radioastronomie, the European Southern Observatory, and the Onsala Space Observatory. } (APEX) 12-m telescope \citep{gun06} in service mode and were carried out in October 7, 9, 11 of 2016, June 21 and September 24-25 of 2017. The APEX Swedish Heterodyne Facility Instrument (SHeFI, Vassilev et al. 2008) band 2 receiver (267~GHz $-$ 378~GHz) was used as a frontend and the eXtended bandwidth Fast Fourier Transform Spectrometer (XFFTS, 2.5~GHz bandwidth and 32768 spectral channels) was used as a backend. The receiver was tuned to the $^{12}$CO~(3$-$2) transition frequency for the $^{12}$CO~(3$-$2) observations and to 329.960~GHz to allow the simultaneous observation of the $^{13}$CO~(3$-$2) and C$^{18}$O~(3$-$2) transition lines. At these frequencies, the beam size ($\theta_{beam}$) and main beam efficiency ($\eta_{MB}$) of the telescope are 19\farcs2 and 0.73, respectively. The Precipitable Water Vapor (PWV) measuring the weather conditions during the observations ranged from 0.9 to 1.6~mm. The observations, performed with the raster mode, consist of 121 ON observations distributed as a 11$\times$11 pixels ($\sim$105\arcsec$\times$105\arcsec) mosaic map separated by half a beam with a OFF reference observed after each 3 ON observations in position switching mode. The observed maps are centered on (258.03625$^{\circ}$,$-$38.51319$^{\circ}$) and (258.17625$^{\circ}$,$-$38.45017$^{\circ}$), referred as region A and region B, respectively (see Fig.~\ref{fig:RCW120_APEX_mapping}). The integration time, excluding overheads, was 3.1 hours for region A in $^{12}$CO~(3$-$2), 3 hours for region A in $^{13}$CO~(3$-$2)/C$^{18}$O~(3$-$2), 1.5 hours for region B in $^{12}$CO~(3$-$2) and 2.8 hours for region B in $^{13}$CO~(3$-$2)/C$^{18}$O~(3$-$2). After the first cycle of observations (2016B), the priority was given to region B in $^{13}$CO~(3$-$2)/C$^{18}$O~(3$-$2) due to the low abundance of these isotopologue compared to $^{12}$CO~(3$-$2). 

Pointings and calibrations were achieved by observing NGC~6302, Mars, RT-Sco and Saturn. The data were further reduced using the CLASS package of the GILDAS software\footnote{\url{http://www.iram.fr/IRAMFR/GILDAS}}. Baselines modelled as 1$^{\rm{st}}$ to 3$^{\rm{rd}}$ order polynomials, depending on the observations, were subtracted. The $^{13}$CO~(3$-$2) and C$^{18}$O~(3$-$2) observations were then tuned to their transition frequency. The \textit{table} and \textit{xy\_map} routines were used to combine the data and create the lmv cubes which were converted into FITS files using the \textit{fits} routine of the VECTOR package. Finally, the cubes were resampled with a pixel size of 9.5\arcsec to the same center and the same size in order to have uniform observations. We ended up with 6 spectral cubes with a resolution of 0.07~km~s$^{-1}$ (76~kHz) and a rms of 0.24~K, 0.54~K, 0.89~K, 0.47~K, 0.31~K and 0.37~K for region A $^{12}$CO~(3$-$2), $^{13}$CO~(3$-$2), C$^{18}$O~(3$-$2) and region B, respectively.

\section{Analysis}

\subsection{Spatial and velocity distribution of CO~(3$-$2), $^{13}$CO~(3$-$2) and C$^{18}$O~(3$-$2)}\label{subsect:spatial_distrib}

\begin{figure*}[t]
  \centering
  \includegraphics[width=.5\textwidth]{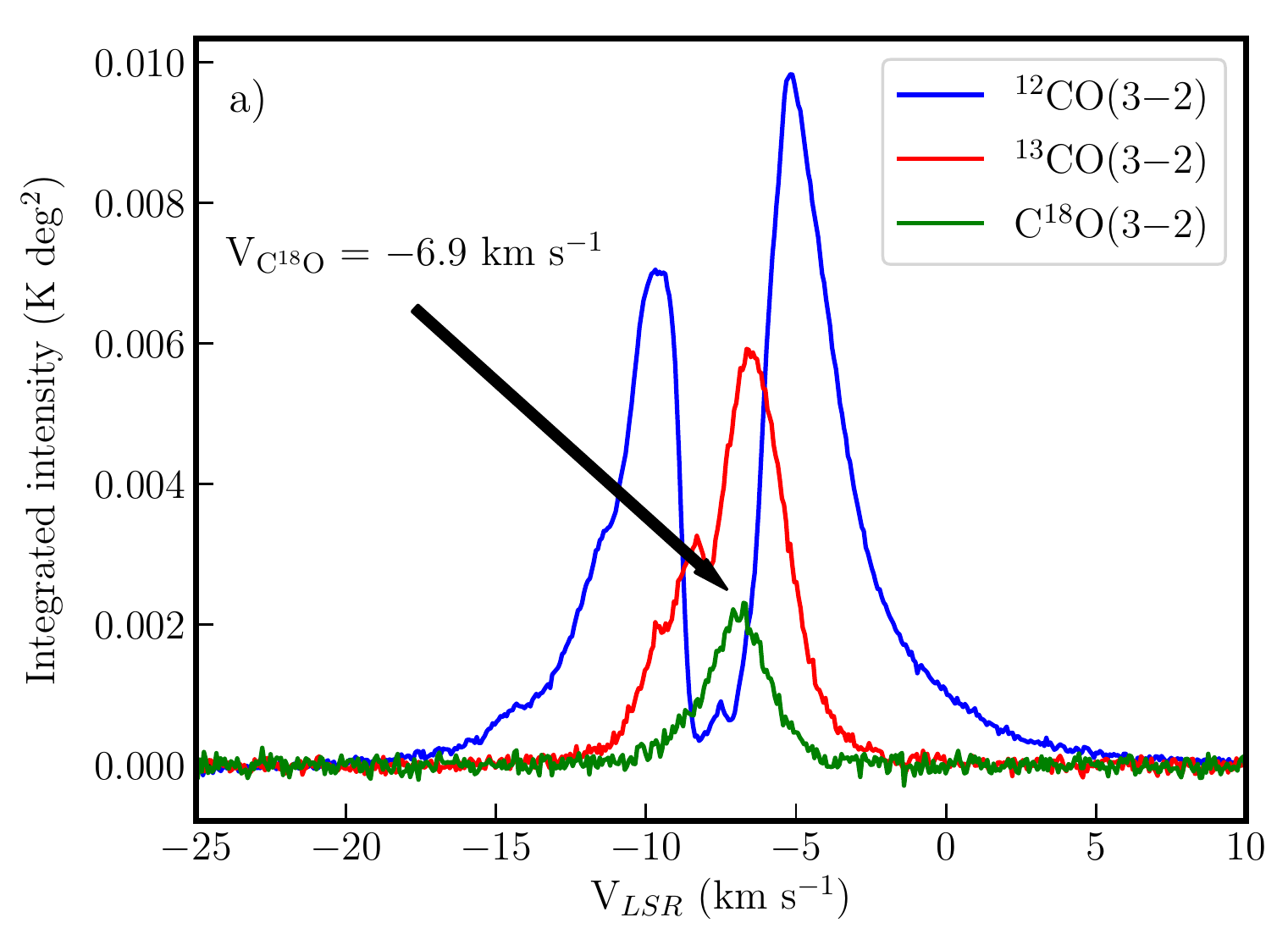}\includegraphics[width=.5\textwidth]{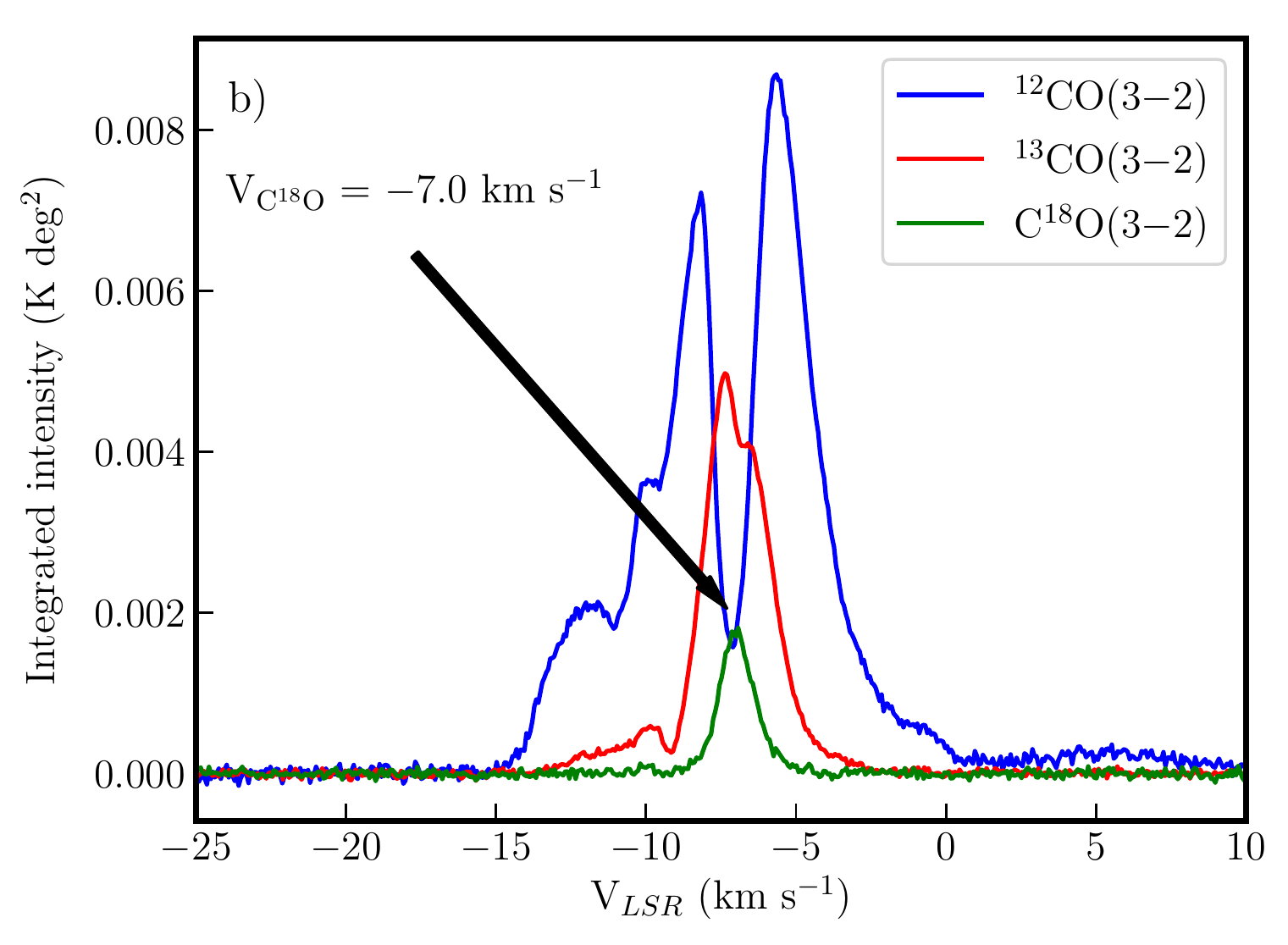}  
  \caption{Spatially integrated intensity of $^{12}$CO (in blue), $^{13}$CO (in red) and C$^{18}$O (in green) toward region A (a) and region B (b)}
  \label{fig:Average_isoCO}
\end{figure*}

\begin{figure*}[t]
  \centering    
  \includegraphics[width=1\textwidth]{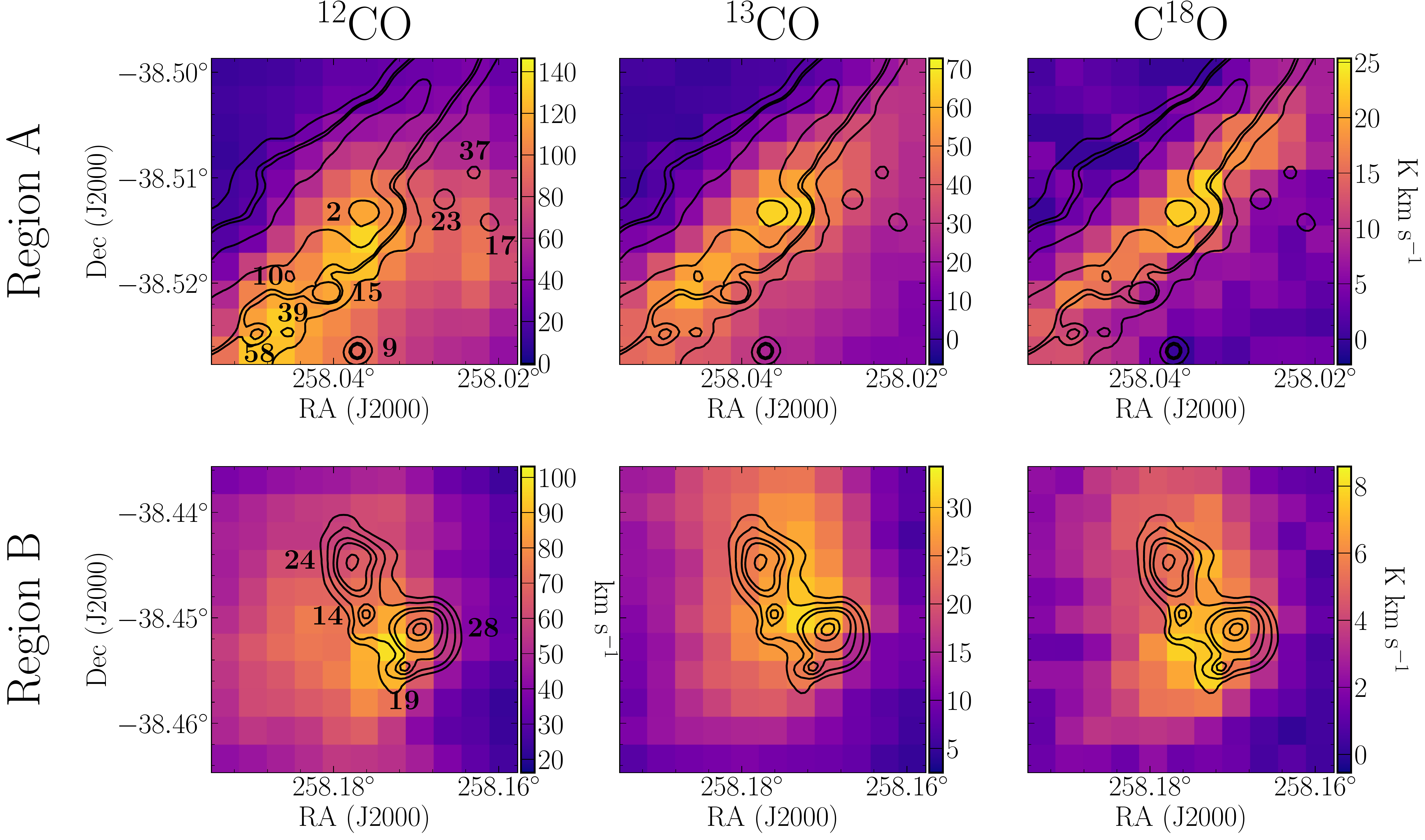}  
  \caption{Velocity integrated intensity of the $^{12}$CO~(3$-$2) (first column), $^{13}$CO~(3$-$2) (second column) and C$^{18}$O~(3$-$2) (third column) for region A (top) and region B (bottom). The velocity range used for integration is $-20.1<V_{\rm{LSR}}<8.2$~km~s$^{-1}$, $-15.0<V_{\rm{LSR}}<3.1$~km~s$^{-1}$ and $-12.0<V_{\rm{LSR}}<-2.3$~km~s$^{-1}$ for $^{12}$CO~(3$-$2), $^{13}$CO~(3$-$2) and C$^{18}$O~(3$-$2), respectively. Contours represent the emission at 70~$\mu$m from \textit{Herschel} and the cores (labels are slightly shifted from the true position of the cores) which were extracted in \citet{fig17}.}
  \label{fig:int_z}
\end{figure*}

The velocity distribution of the spatially integrated emission and the velocity integrated maps for the three isotopologues and both regions are shown in Figs.~\ref{fig:Average_isoCO},\ref{fig:int_z}. Due to the abundance, the integrated intensity is decreasing from $^{12}$CO to C$^{18}$O with a ratio to the $^{12}$CO peak of 1, 0.51 and 0.20 for region A, and 1, 0.51 and 0.18 for region B. The $^{12}$CO and $^{13}$CO double peaks behaviour could either indicates the presence of two clumps or a strong self-absorption. Since this is not observed in the optically thin C$^{18}$O transition, the double peak feature is a signature of self-absorption, clearly observed in $^{12}$CO due to high optical depth ($\tau$). It was also observed by \citet{and15} in $^{12}$~CO($1-0$) using MOPRA observations. The optical thickness of $^{12}$CO ($\tau_{12}$) can be estimated by comparing the intensity ratio of $^{12}$CO to $^{13}$CO to their relative abundance \citep{haw13}. Assuming a $^{12}$CO and $^{13}$CO abundance of 8$\times$10$^{-5}$ and 2.7$\times$10$^{-6}$, respectively \citep{mag88,pin08}, the abundance ratio is $\sim$30, much higher than the average line intensity ratio of 2.4, indicating a high optical depth for $^{12}$CO. The $^{13}$CO presents some self-absorption features, especially toward region A, in agreement with the fact that clump 1 is the densest of RCW~120.

The gaussian fit of C$^{18}$O indicates that this molecular line is centred at $-$6.9~km~s$^{-1}$ for region A and $-$7.0~km~s$^{-1}$ for region B, respectively. RCW~120 having a V$_{LSR}$ of $-$7 to $-8$~km~s$^{-1}$, these two emissions are associated with its PDR.\\
In the southern-east part of region A, two features are observed. A small peak is detected only in $^{12}$CO at $\sim -$15~km~s$^{-1}$ (white pixels on Fig.~\ref{fig:spectra_fig_int_z}) with a $T_{MB}$ less than 5~K. This feature may be too weak to be detected in other isotopologues. It could be due to line of sight contamination as it does not appear to be related to the PDR or the ionized region. The other feature can be seen at $-$12.5~km~s$^{-1}$ and located in the north-western part of the region, where the ionized gas is, and being weaker toward the PDR. Its location on the blue side part of the spectrum may represents the $^{12}$CO gas moving toward us due to the ionization pressure. They can only be weakly seen on the spatially averaged profile (Fig.~\ref{fig:Average_isoCO}) at $\sim -14.5$ and $\sim -12$~km~s$^{-1}$. Toward region B, we note the non gaussian profile with asymmetry on the blue part. Except from the main one, three other components can be seen at $\sim-$12, $-$10 and $-$1~km~s$^{-1}$, discussed in Sect.~\ref{sect:rdi}.\\
 
\begin{figure*}
  \centering    
  \includegraphics[width=1\textwidth]{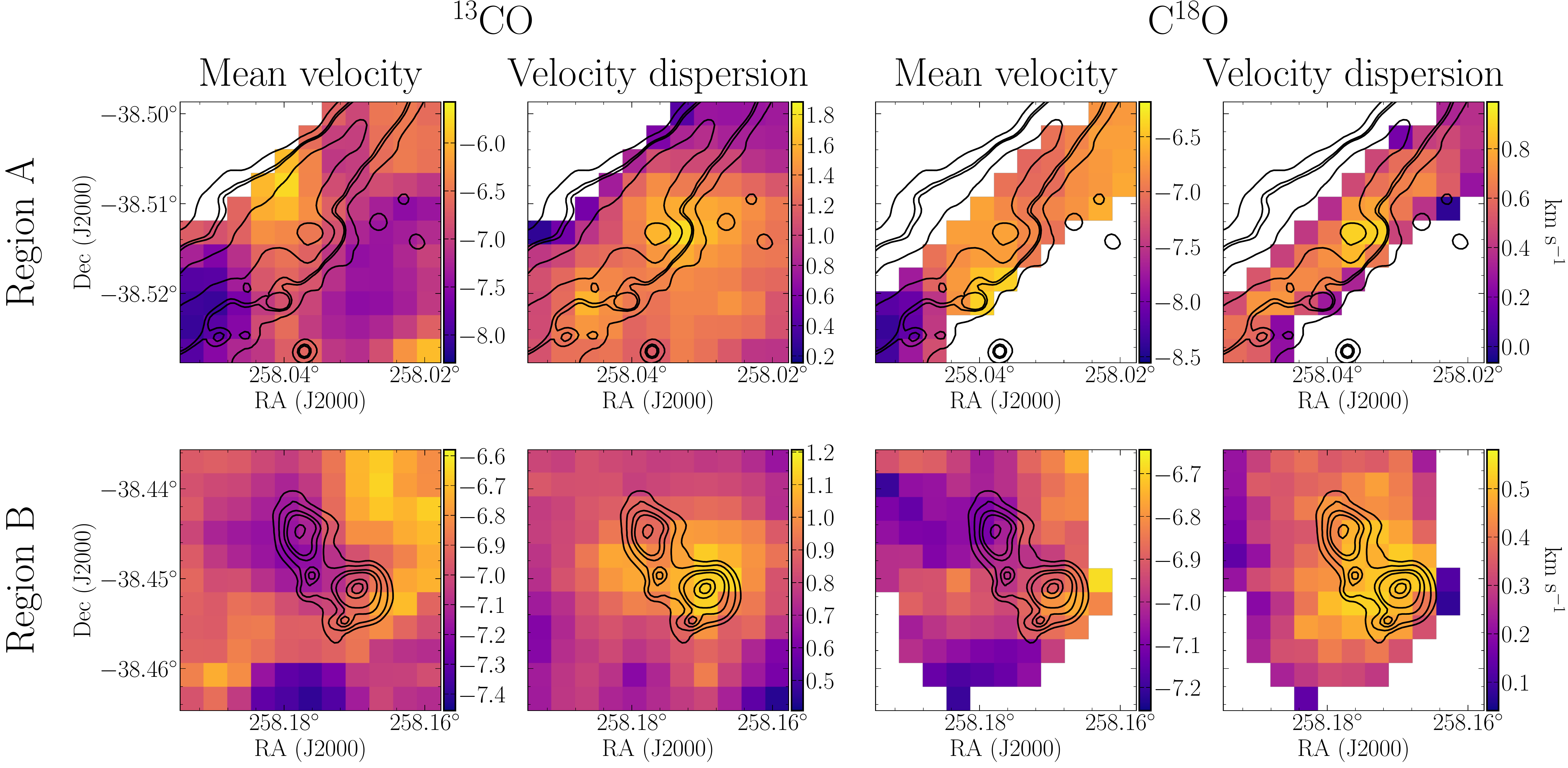}
  \caption{Mean velocity and velocity dispersion maps of $^{13}$CO~(3$-$2) (first and second columns) and C$^{18}$O~(3$-$2) (third and fourth columns) towards region A (top) and B (bottom). The integration velocity and contour levels are the same as for Fig.~\ref{fig:int_z}.}
  \label{fig:mean_velo_sigma}
\end{figure*}

We constructed the mean velocity and velocity dispersion maps\footnote{The mean velocity and velocity dispersion are computed following M$_{1}$=$\int_{}^{}T_{MB}VdV/\int_{}^{}T_{MB}dV$ and M$_{2}$=$\int_{}^{}T_{MB}(V-M_{1})dV/\int_{}^{}T_{MB}dV$}, also known as first and second moment maps, of the $^{13}$CO and C$^{18}$O and a clip at 3$\sigma$ (Fig.~\ref{fig:mean_velo_sigma}). The mean velocity of the $^{13}$CO and C$^{18}$O in region A ranges from $-$8.3 to $-$5.5~km~s$^{-1}$ with an average of $-$7~km~s$^{-1}$. While a range of different velocities is observed, the interval is too small and no clear gradients or particular features are observed. The average velocity dispersion of 0.7~km~s$^{-1}$ for C$^{18}$O, in good agreement with the velocity dispersion of 0.8~km~s$^{-1}$ found towards the PDR of RCW~120 with MOPRA CO~(1$-$0) data \citep{and15}. It also seems in agreement for $^{13}$CO with an average of 1.4~km~s$^{-1}$ (see their Fig.~9). On the velocity dispersion maps, the center of region A stands out with a velocity dispersion of 2.6 and 1.1~km~s$^{-1}$ for $^{13}$CO and C$^{18}$O, respectively, and shows the increase of linewidth, through turbulence and thermal contributions from the Class~0 \textit{Herschel} source 2.\\
In region B, the mean velocity of the $^{13}$CO and C$^{18}$O ranges from $-$7.2 to $-$6.6~km~s$^{-1}$ with an equal average of $-$6.9~km~s$^{-1}$. In $^{13}$CO, the mean velocity does not seem to be distributed randomly in the map with lower mean velocity toward the H{\,\sc{ii}} region and higher toward the clump but, as for region A, the range of velocity is too small to provide any meaningful conclusion. Higher spatial and spectral resolution observations are needed to study gas motions in these two regions. The velocity dispersions of 0.8 and 0.4~km~s$^{-1}$, in average, are also consistent with the MOPRA observations. Moreover, there are higher where the evolved low-mass sources are located.\\
Due to the temperature difference between the most massive core of RCW~120 (labelled as core 2, see \citealt{fig17} and Fig.~\ref{fig:int_z}) with $T_{dust}=20.3$~K and the YSOs in region B with $T_{dust}=34.1$~K, turbulence toward region A should be higher in order to explain the higher velocity dispersion. This is in agreement with the high-mass star formation occurring in this core \citep{fig18} compared to the low-mass YSOs present in region B.

\subsection{Self-absorption correction}\label{subsect:absorbtion_correction}

\begin{figure}
  \centering
\includegraphics[width=\linewidth]{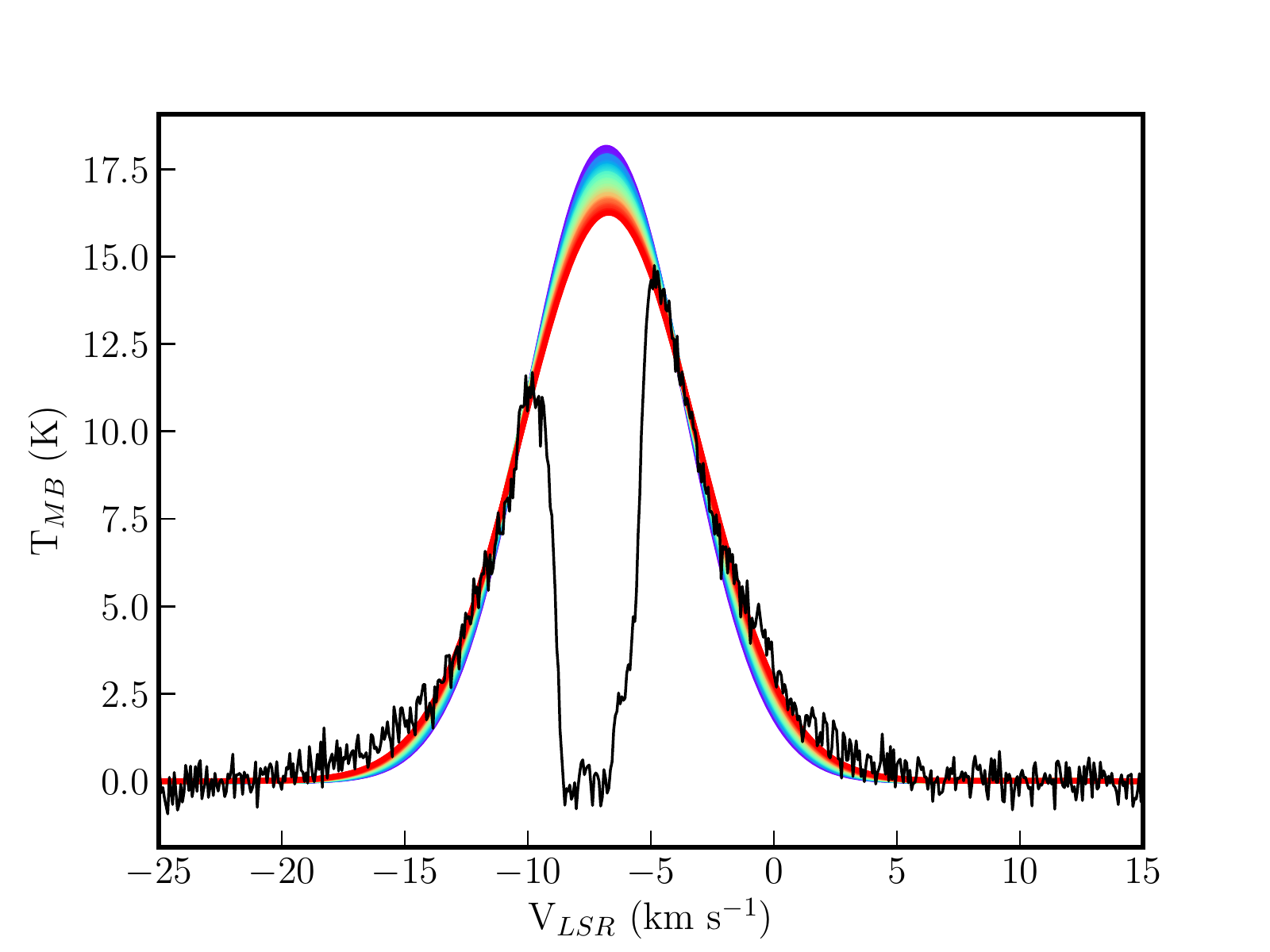}
  \caption{$^{12}$CO spectrum in region A. The purple curve corresponds to the fit using the upper part of the shoulders while the red curve is obtained using the whole spectrum except the self-absorption feature.}
  \label{fig:self_absorption}
\end{figure}

The self-absorption feature seen in $^{12}$CO and $^{13}$CO (see Fig.~\ref{fig:Average_isoCO}) is responsible for the intensity loss at the $V_{LSR}$ of the region ($-$7~km~s$^{-1}$) resulting in a lower $T_{MB}$ than expected. To correct the spectra for self-absorption, we performed iterative gaussian fittings using the shoulder of the profile. We began with the upper part of the profile shoulders and iteratively perform fittings by increasing their length until reaching the end of the profile. Secondary peaks are removed during the fitting. As seen on Fig.~\ref{fig:self_absorption}, when using a smaller length (bluer color), the peak tends to be higher compared to larger shoulders' length (redder color). The uncertainty on the peak value was computed as the standard deviation of the different peak values while it was computed as the 1$\sigma$ uncertainty if no self-absorption was observed. Hence, each spectrum could be fitted with a single gaussian and a peak main brightness temperature could be derived for each of them ($T_{\rm{MB}}^{12}$, $T_{\rm{MB}}^{13}$ and $T_{\rm{MB}}^{18}$).  As self-absorption decreases, either going from $^{12}$CO to C$^{18}$O or from region A to region B where CO is less abundant, the uncertainties on $T_{MB}$ decrease. Toward region B, 100\% and 34\% of the spectra are corrected for self-absorption while 88\% and 23\% for region A, for $^{12}$CO and $^{13}$CO. In the following, all the peak values are corrected from self-absorption.

\subsection{Physical properties of the clumps}   

The solution of the radiative transfer equation can be used to derive several physical properties of the clumps such as the excitation temperature $T_{ex}$, the optical depth $\tau$ or the column density $N$. Assuming that the medium is uniform, the background subtracted solution of the radiative transfer equation can be written explicitly using the Planck law and rearranged to obtain $T_{ex}$ as a function of $T_{MB}$, the CMB temperature $T_{CMB}$ and $\tau$ \citep{roh96,haw13}:

\begin{equation}\label{eq:solution_transfert}
T_{ex}=T_{\nu}\left[ {\rm{ln}} \left( 1+T_{\nu}\frac{1-e^{-\tau}}{T_{MB}+T_{\nu}(1-e^{-\tau})e^{\frac{-T_{\nu}}{T_{CMB}}}} \right) \right]^{-1}
\end{equation}

where $T_{\nu}=h\nu/k_B$, $k_{B}$ is the Boltzmann constant, $h$ is the Planck constant and $\nu$ is the frequency transition of the line considered. The parameters of Eq.~\ref{eq:solution_transfert} can be found in Tab.~\ref{tab:para_LTE} for the $J=3-2$ transition. As shown in Sect.~\ref{subsect:spatial_distrib}, $^{12}$CO is optically thick so $T_{ex}$ can be simplified considering $\tau \rightarrow \infty$.

\begin{figure*}
  \centering
  \includegraphics[width=1\textwidth]{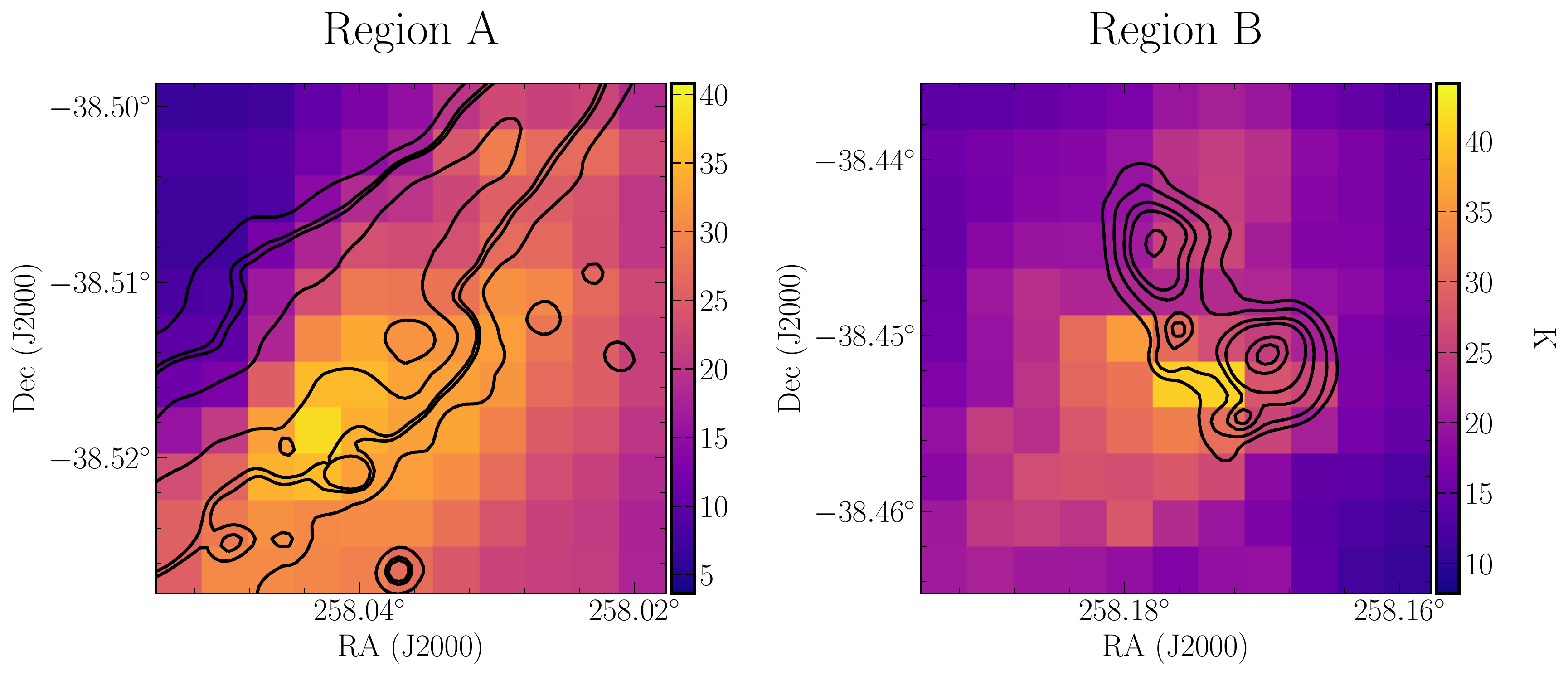}
  \caption{$T_{ex}$ toward region A (left) and region B (right) obtained using Eq.~\ref{eq:solution_transfert}.}
  \label{fig:A_Tex_tau}
\end{figure*}

Toward region A, $T_{ex}$ is minimal inside the H{\,\sc{ii}} region (6.7~K) where the $^{12}$CO is less abundant, increases along the PDR where star formation is observed, with the highest value (38.3~K) towards the center where the \textit{Herschel} source 2 is located, with an average of 22.8~K. Toward region B, the excitation temperature is also minimal inside the H{\,\sc{ii}} region (10.5~K) and increases, up to 40.9~K, towards the center of the clump, where evolved low-mass stars are observed, with an average of 20.7~K. These values agrees with $T_{ex}=23.4$~K from \citet{and15} and are similar to those toward the mid-infrared bubble S~44 \citep{koh18} where ranges of 8$-$13~K and 8$-$25~K were found. The locations of the $T_{ex}$ high values (see Fig.~\ref{fig:A_Tex_tau}) are associated with the \textit{Herschel} sources which, together with the FUV radiation from the H{\,\sc{ii}} region, is another important source of heating.\\

By inverting Eq.~\ref{eq:solution_transfert} and assuming that Local Thermodynamical Equilibrium (LTE) holds, $\tau _{13}$ and $\tau _{18}$ can be computed through Eq.~\ref{eq:tau_LTE}.

\begin{equation}\label{eq:tau_LTE}
\tau=-{\rm{ln}}\left[1-\frac{T_{\rm{MB}}}{T_{\nu}(\frac{1}{e^{T_{\nu}/T_{ex}}-1}-e^{-T_{\nu}/T_{\rm{CMB}}})}\right]
\end{equation}

Towards region A, the optical depth of $^{13}$CO and C$^{18}$O follows the same behaviour than $T_{ex}$ with low values outwards the PDR (around 0.2 and 0.1 inside the H{\,\sc{ii}} region) and higher values (up to 2.3 and 0.7) towards the PDR. In average, $^{13}$CO has a larger optical depth compared to C$^{18}$O, the former being moderately thick along the PDR and the latter optically thin everywhere. In region B, the values of $\tau_{13}$ and $\tau_{18}$ ranges from 0.2 and 0.1, inside the H{\,\sc{ii}} region, to 2.7 and 0.4, respectively with a similar average compared to region A. We have to note that the highest values of $\tau$ are always found at the edges of the map and we cannot exclude that the resampling process lowered the quality of the border of the map. If these high values are excluded, $^{13}$CO is moderately thick with $\tau_{13}$ reaching values up to 1.5.\\

By studying a sample of bright rim clouds (BRCs), \citet{mor09} showed that $T^{18}_{ex}$ can be significantly different from $T^{12}_{ex}$ since C$^{18}$O probes the interior of the clump due to its low optical thickness compared to $^{12}$CO which is mostly representative of the surface of the clump \citep{tak19}. Using Eq.~\ref{eq:solution_transfert}, an estimation of $T_{ex}$ for C$^{18}$O can be computed by using the average $\tau_{18}$ found with Eq.~\ref{eq:tau_LTE}. To make a comparison with $T^{12}_{ex}$, we only take into account the $^{12}$CO pixels where C$^{18}$O is detected. In average, $T^{12}_{ex}$ is higher than $T^{18}_{ex}$ as observed but with a lower temperature difference (2$-$4~K) than in \citet{mor09}. Stars and protostars located inside region A and B could explained this rise of $T^{18}_{ex}$ as clumps are not quiescent.\\

The column density of $^{13}$CO and C$^{18}$O can be obtained with:

\begin{equation}\label{eq:column_density_LTE}
N=\left(\frac{3k_B}{8\pi^3B\mu^2}\right)\left(\frac{e^{hBJ_l(J_l+1)/k_BT_{ex}}}{J_l+1}\right)\left(\frac{T_{ex}+hB/3k_B}{1-e^{h\nu/k_BT_{ex}}}\right)\int \tau dv
\end{equation}

\begin{equation}
\int \tau dv=\frac{1}{T_{\nu}(\frac{1}{e^{T_{\nu}/T_{ex}}-1}-e^{-T_{\nu}/T_{\rm{CMB}}})}\frac{\tau}{1-e^{-\tau}}\int T_{\rm{MB}}dv
\end{equation}

Towards region A, N($^{13}$CO) follows the PDR of RCW~120 with the maximum of 5.5$\times$10$^{16}$~cm$^{-2}$ found towards the most massive core of the region. The N(C$^{18}$O) distribution also follows the PDR but with a smaller variation, a maximum of 1.4$\times 10^{16}$~cm$^{-2}$ toward the massive core and a decrease outwards. Towards region B, the N($^{13}$CO) and N(C$^{18}$O) distributions are well correlated with the dust continuum emission with a value toward the center of 2.7$\times 10^{16}$ and 4.1$\times 10^{15}$~cm$^{-2}$. Since the column density calculation depends on $\tau$, a similar issue arises toward the edge of the map, mainly seen for N($^{13}$CO) in both region.
The global values of column density found toward this region are in agreement with other star-forming regions \citep{shi14,par18,vaz19}. Inside the H{\,\sc{ii}} region, no $^{13}$CO~(3$-$2) and C$^{18}$O~(3$-$2) are detected. Given the rms of these observations, it translates into an upper limit for the column density of 1.5$\times$10$^{15}$~cm$^{-2}$ and 7.5$\times$10$^{13}$~cm$^{-2}$ for $^{13}$CO and C$^{18}$O, respectively.\\

 \begin{table}
\caption{Parameters used to compute $T_{ex}$, $\tau$ and $N$}            
\label{tab:para_LTE}      
\centering                          
\begin{tabular}{c|c|c|c|c}
\hline
 \hline
CO~(3$-$2) & $\nu$ &$T_{\nu}$ & exp$(-T_{\nu}/T_{CMB})$ & B   \\
$J_l=2$ & (GHz) & (K) &  & (GHz)  \\
\hline
 \hline
$^{12}$CO & 345.795 & 16.6  & 0.0022  & 57.635  \\
$^{13}$CO & 330.587 &15.9 & 0.0029  & 55.101  \\
C$^{18}$O & 329.330 &15.8 & 0.0030 & 54.891  \\
\hline
 \hline
\end{tabular}
\end{table}

\begin{figure*}[t!]
  \centering
\includegraphics[width=1.\textwidth]{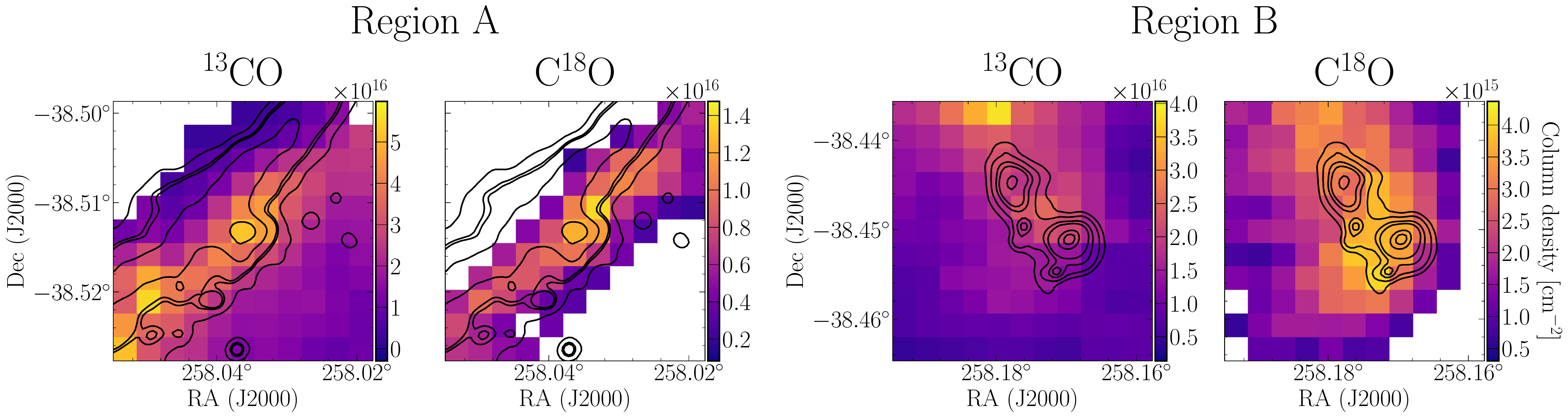}
  \caption{Column density of $^{13}$CO and C$^{18}$O for region A (first and second columns) and toward region B (third and fourth columns).}
  \label{fig:column_density_LTE}
\end{figure*}

\begin{table*}
\caption{$T_{ex}$ from $^{12}$CO and C$^{18}$O, optical depth and column density of $^{13}$CO and C$^{18}$O} 
\label{tab:T_LTE_RADEX}      
\centering                         
\begin{tabular}{c|c|cc|cccc}
\hline\hline
&	&	 $T_{ex}^{12}$  	&	 $T_{ex}^{18}$ 	&	  $ \tau_{13}^{\rm{LTE}}$ 	&	 $ \tau_{18}^{\rm{LTE}}$ 	&	  $ N(^{13}$CO$)^{\rm{LTE}}$ 	&	 $ N($C$^{18}$O$)^{\rm{LTE}}$ \\
&	&	 (K)  	&	 (K) 	&	 	&	 	&	 (cm$^{-2}$) 	&	 (cm$^{-2}$) \\
	\hline
&Minimum 	&	6.7 (18.7)	&	14.8	&	0.2	&	0.1	&	  2.1$\times 10^{15}$  	&	  1.9$\times 10^{14}$ \\
Region A &Maximum 	&	38.3 (38.3)	&	39.1	&	2.3	&	0.7	&	  5.5$\times 10^{16}$ 	&	  1.4$\times 10^{16}$ \\
&Mean 	&	22.8 (28.3)	&	24.8	&	0.7	&	0.3	&	  2.2$\times 10^{16}$ 	&	  6.5$\times 10^{15}$ \\
\hline																				
&Minimum  	&	10.5 (13.9)	&	9.7	&	0.2	&	0.1	&	  4.4$\times 10^{15}$ 	&	  6.2$\times 10^{14}$  \\
Region B&Maximum  	&	40.9 (40.9)	&	34.7	&	2.7	&	0.4	&	  3.8$\times 10^{16}$ 	&	  4.1$\times 10^{15}$ \\
&Mean 	&	20.7 (21.8)	&	19.9	&	0.7	&	0.2	&	  1.3$\times 10^{16}$ 	&	  2.1$\times 10^{15}$  \\
\hline
\hline
\end{tabular}
\end{table*}

\subsection{Outflows toward RCW~120}\label{subsect:outflow_extraction}
\subsubsection{Extraction of the wings}

\begin{figure*}[t]
  \centering
  \includegraphics[width=.5\textwidth]{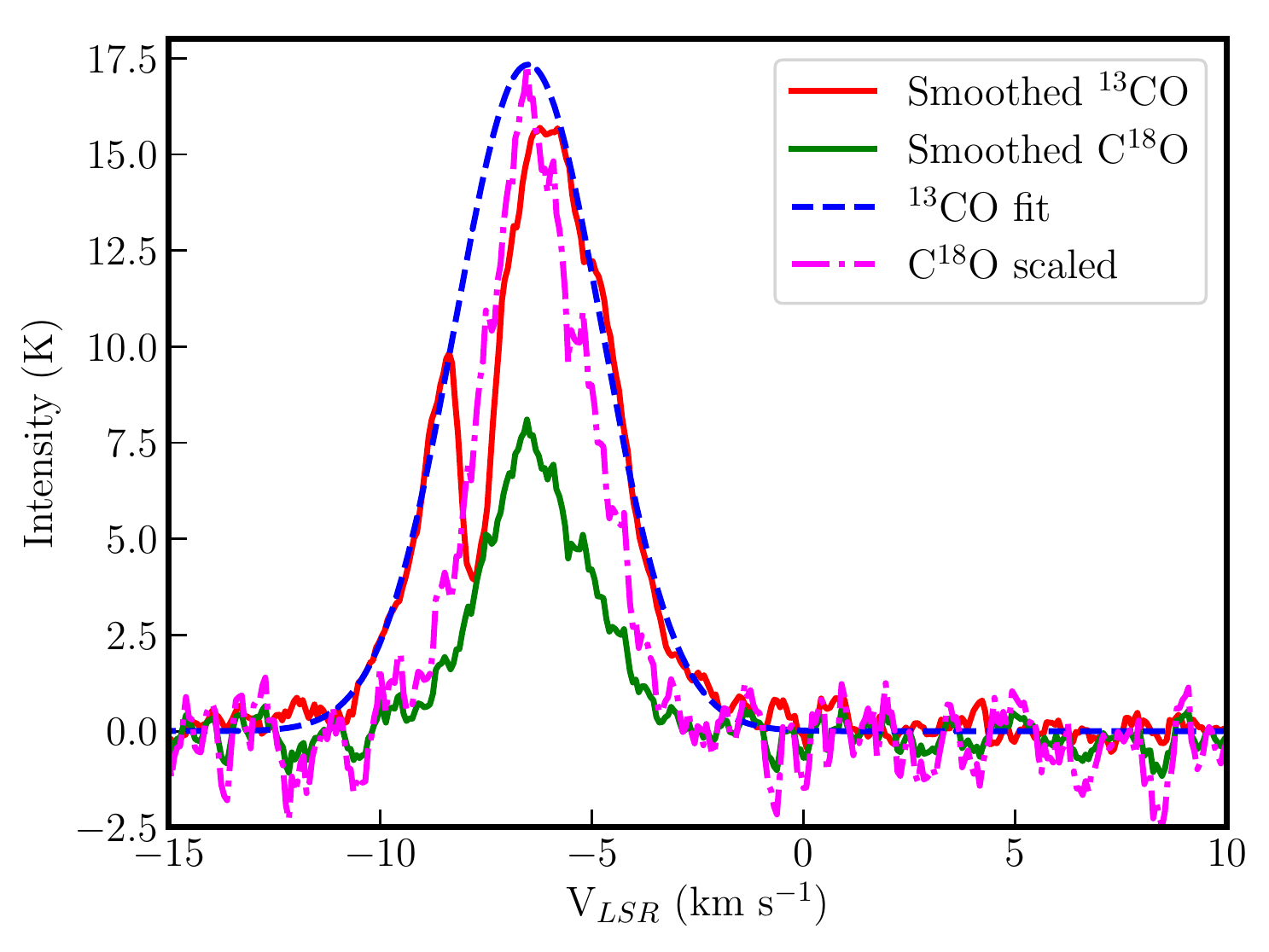}\includegraphics[width=.5\textwidth]{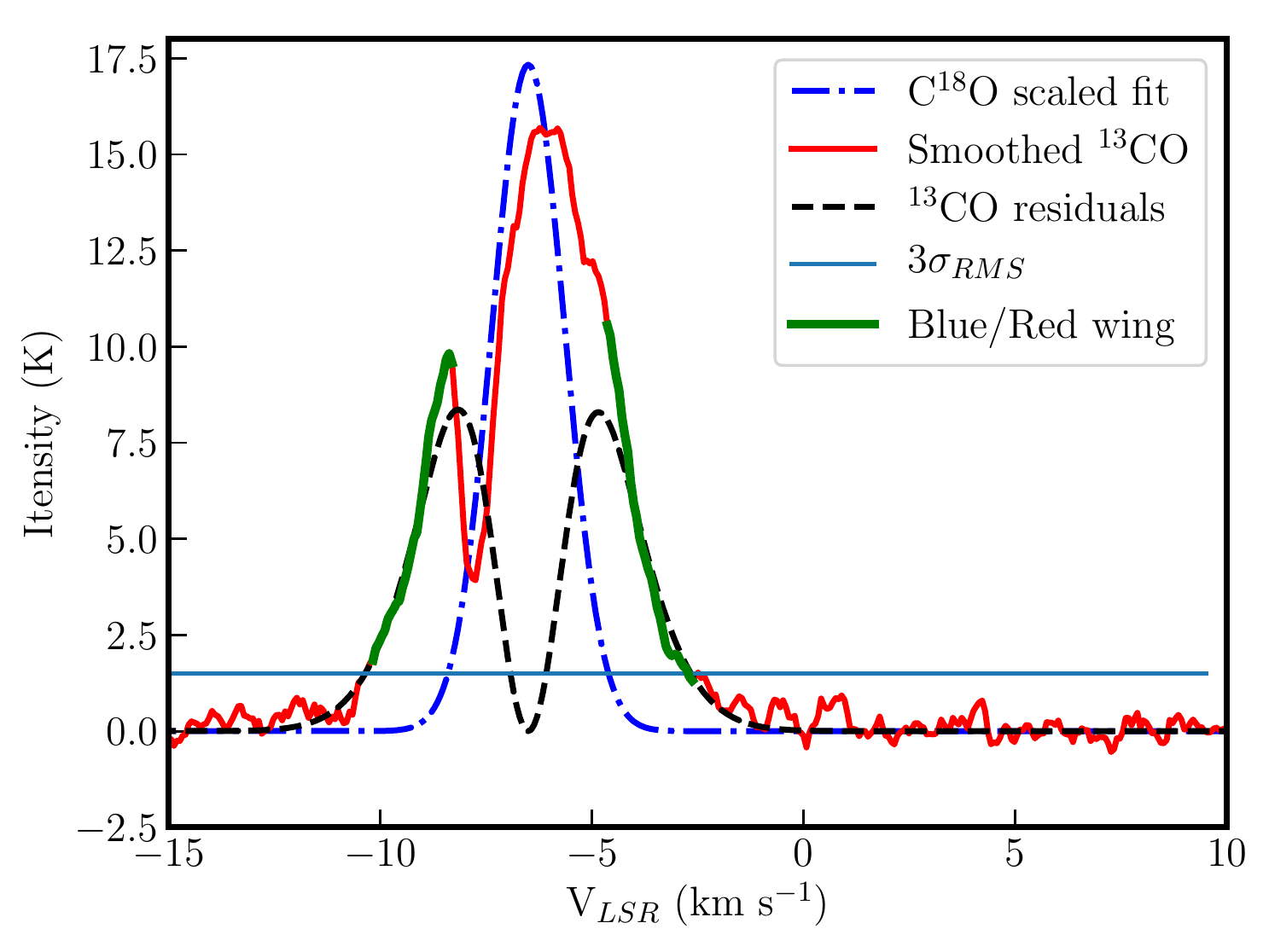} 
  \caption{Application of the method described in Sect.~\ref{subsect:outflow_extraction} to the source 2 of region A. Left: $^{13}$CO and C$^{18}$O spectra convolved with a one dimensional box of 5 pixels (continuous red and green lines), the gaussian fit of the $^{13}$CO to correct for the self-absorption (dashed blue line) and the C$^{18}$O spectrum scaled to the peak of the $^{13}$CO gaussian fit (dashed-point magenta line). Right: gaussian fit to the scaled C$^{18}$O spectrum (dashed-point blue line), $^{13}$CO residual (dashed black line) resulting from the subtraction of the scaled C$^{18}$O gaussian from the $^{13}$CO gaussian and the blue and red wings (green line).}
  \label{fig:Region_AB_Wings}
\end{figure*}

\begin{figure}[b]
  \centering
  \includegraphics[width=.5\textwidth]{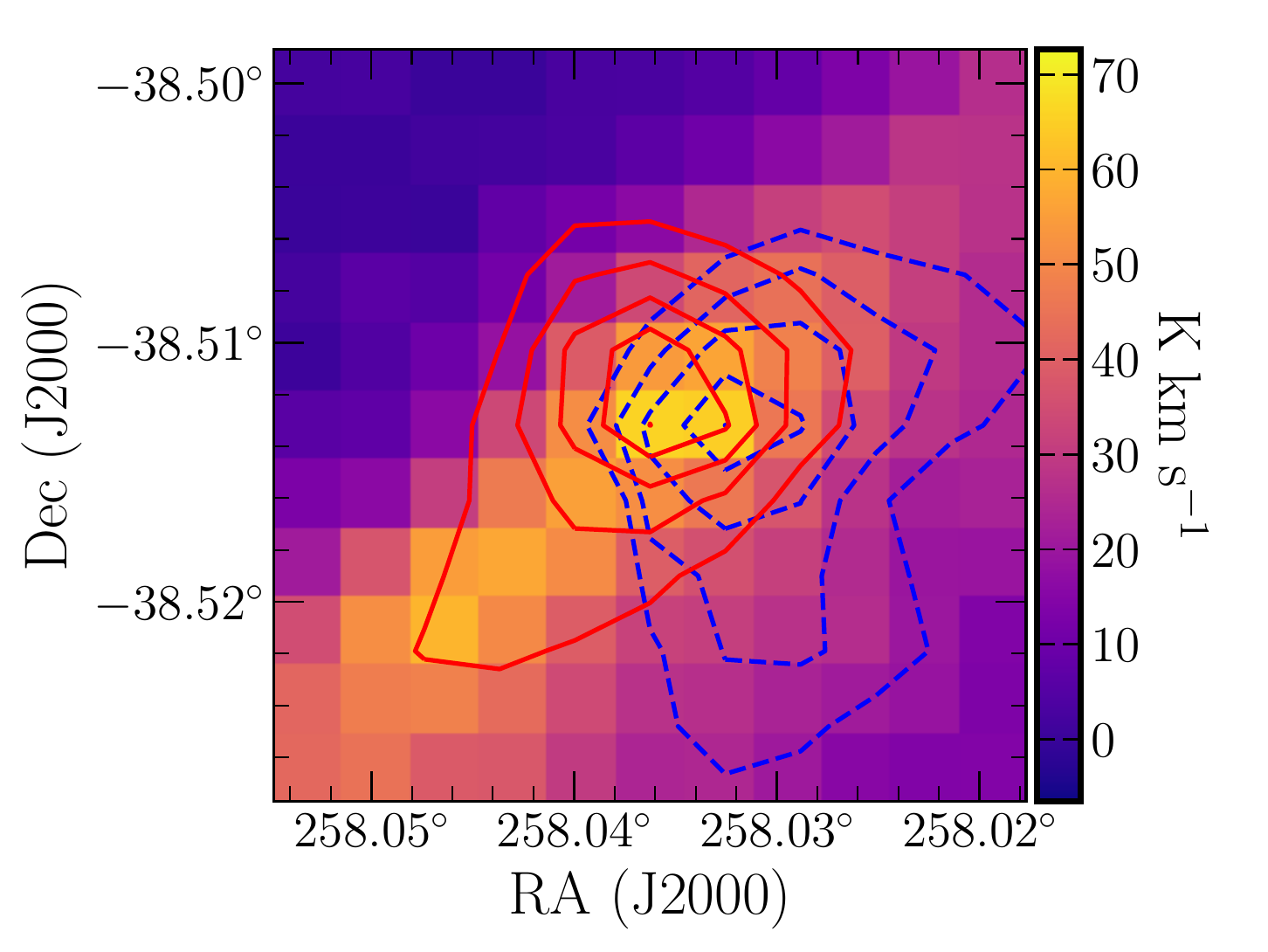}%
  \caption{Outflow contours in region A for the red and blue wings (continuous red and dashed blue contours) overlaid on the $^{13}$CO~(3$-$2) velocity integrated intensity image.}
  \label{fig:A13CO_z_Wings}
\end{figure}

During the phase of high-mass star formation, one of the solution proposed to overcome the radiation pressure problem \citep{wol87} was the growth of the star via an accretion disk \citep{jij96} such as in the formation of low-mass stars \citep{cof08}. For the angular momentum to be conserved, high-mass stars formed by disk accretion must radiate this excess of momentum via outflows. Observations showed that high-mass star formation are associated with outflows \citep{arc07,mau15,mcl18}. Since they release more momentum, energy and spawn a larger distance, they are more easily detectable compared to the low-mass ones. During high-mass star formation, they developed during the hot-core stage preceding the UCH{\,\sc{ii}} phase, also associated with the 6.67~GHz class~II methanol maser \citep{cas13}. The detection of outflows is mostly done through the identification of wings on spectral profiles but other indirect tracers of outflow exists such as SiO. \\
Toward region A, the source 2 observed with \textit{Herschel} is a massive source where high-mass star formation is observed through the hot core phase \citep{fig18}. Several molecular transitions (MALT90 survey) such as tracers of hot core (CH$_3$CN and HC$_3$N) as well as a tracer of shock and outflow (SiO) are observed toward the source 2, strongly indicating that an outflow is present. The profile is broad \citep{kir19} and synonym of dynamics toward this core. SiO is also detected toward source 39. As it is placed at the edge of the map, we will not extract the outflow since the edges are problematic due to the resampling process. However, it may contaminate the properties of the outflow at the location of source 2. Toward Region B where low-mass sources are observed, no wings and no outflow tracers are detected. Sources are classified as Class~I, II or Ae/Be stars \citep{deh09,fig17} and due to the more evolved stage of these sources, the outflows, if any, should be weaker compared to region A.\\
To extract the outflow wings from our spectra, we employed the method of \citet{dev14} used in the framework of methanol masers as counterpart of the hot core stage, also used in \citet{yan18} toward ATLASGAL clumps. The procedure to extract the wings of the spectrum is visually explained on Fig.~\ref{fig:Region_AB_Wings} and the basic idea is to retrieve the part of the $^{13}$CO spectrum which is broader than the C$^{18}$O spectrum. Firstly, the C$^{18}$O spectrum is scaled to the peak of the $^{13}$CO spectrum which is then fitted with a gaussian. In order to remove high velocity features \citep{van07} and correctly fit the spectrum peak, we first performed a fit of the whole spectrum and we iteratively fitted the spectrum by reducing the shoulders length, pixel to pixel (see Fig.~\ref{fig:Region_AB_Wings} left). This modelled scaled C$^{18}$O spectrum is then subtracted from the observed $^{13}$CO spectrum to obtain the $^{13}$CO residuals. The blue and red wings are defined as the part of the $^{13}$CO spectrum corresponding to the $^{13}$CO residuals above 3$\sigma$ and below the two maxima of the residuals or, said in another way, the part of the $^{13}$CO spectrum which is broader than the scaled C$^{18}$O spectrum (see Fig.~\ref{fig:Region_AB_Wings} right). Toward region A, the $^{13}$CO spectrum was corrected for self-absorption, as explained in Sect.~\ref{subsect:absorbtion_correction}. When a gaussian fitting was performed, we first convolved the signal using a one dimensional box of 5 pixels in order to remove the fluctuations. By visual inspection, we checked that this process does not smooth the particular features of our spectra (self-absorption, wings, secondary peaks).\\
The $^{13}$CO spectrum was then integrated in the velocity windows defined by the wings. These wings are presented on Fig.~\ref{fig:A13CO_z_Wings} where the lowest contour is visually chosen as the one enclosing the wings on the integrated map. As it is difficult to differentiate between the background and the emission wings, this leads to a higher uncertainty when deriving the parameters of the outflow as they depend on the wings mass which depends on the lowest contour used.

\subsubsection{Outflow properties}\label{subsubsect:outflow_properties}

The mass of the blue and red wings, represented in Figs.~\ref{fig:Region_AB_Wings}, \ref{fig:A13CO_z_Wings}, are computed by integrating the lobes areas on the N($^{13}$CO) map integrated over the wings velocity ranges. Using the abundance of $^{13}$CO relative to H$_2$ from \textit{Herschel} observations, N($^{13}$CO) is converted into a mass. The total mass of the outflow is obtained by summing the contribution of the blue and red lobes. An estimation of the outflow momentum and energy were obtained by summing each contribution from the velocity channels corresponding to the blue and red wings. The dynamical timescale $t_D$, the mass-loss rate $\dot{M_{out}}$, the mechanical force $F_m$, and luminosity $L_m$ of the outflow were then be computed (see Eqs.~4-9 in \citealt{yan18}).\\
The properties of the outflows in region A can be found in Tab.~\ref{tab:outflows}. When computing the properties of outflows, we have to account for several parameters which increase their uncertainty (see Sect.~4.3 of \citealt{dev14}). One of the most severe is the inclination of the outflow with respect to the line of sight. If this property is unknown, either the corrections are not done, either they are applied assuming a uniform distribution in the sky ($\theta=57.3^{\circ}$). In this paper, we do not correct for the inclination but the reader can refer to Table~4 of \citealt{dev14} and apply the correction factors in case of comparison with other works. The values of the massive core 2 outflow (376~$M_{\odot}$, 856~$L_{\odot}$ \citealt{fig18}) are in relative good agreement with the general values for massive star forming (MSF) clumps derived by \citet{yan18} with a sample of 153 ATLASGAL clumps and by \citet{dev14} from methanol maser associated outflows. Since mass of the wings derived from N($H_2$) might be overestimated, we account for an uncertainty on the outflow properties between a factor 2$-$3 relatively to those in \citet{dev14} and \citet{yan18}.\\

\begin{table}
\caption{Outflow properties}
\label{tab:outflows}      
\centering                          
\begin{tabular}{l|c}
Parameters & Core 2 \\
\hline
\hline
$M_{out}$ ($M_{\odot}$) & 116 \\
$p$ ($M_{\odot}$~km~s$^{-1}$) & 281 \\
$E$ (J) & 7.3$\times 10^{38}$ \\
$t_D$ (yr) & 1.6$\times 10^{4}$  \\
$\dot{M_{out}}$ ($M_{\odot}$~yr$^{-1}$) & 7$\times 10^{-3}$ \\
$F_m$ ($M_{\odot}$~km~s$^{-1}$~yr$^{-1}$) & 1.7$\times 10^{-2}$  \\
$L_m$ ($L_{\odot}$) & 3.7 \\
\hline
\end{tabular}
\end{table}

\subsection{Estimation of the Mach number}

The turbulence in molecular clouds is an important parameter to take into account in star formation as it can counteract the gravity during the gravitational collapse of a core. The solenoidal and compressive modes of turbulence have different impact on star-formation where the latter one is thought to be associated with a higher Star Formation Efficiency (SFE, \citealt{fed12}). The goal here is not to derive the ratio of solenoidal to compressive modes as it has been done in \citet{ork17} but rather to obtain an estimation of the Mach number $M$. The thermal linewidth of the line $c_{T}$ is defined by $\sqrt{k_BT/ \mu m_H}$ where $T$ is taken to be the maximal temperature between $T_{ex}$ (from $^{12}$CO) and $T_{dust}$ (from \textit{Herschel} observations) as recommended by \citet{ork17}. The non-thermal component of the linewidth is $\sigma_{NT}=\sqrt{\sigma^2-\sigma_{T}^2}$ where $\sigma$ is the dispersion of the $^{13}$CO spectra related to the FWHM by $\sigma=FWHM/\sqrt{8\rm{ln}2}$. The Mach number is defined as $M=\sigma_{NT}/c_S$ where $c_s$ is the sound speed. The median for region A and region B is 17 and 7, respectively. In \citet{ork17}, $M$ can goes up to value higher than 20 towards the regions where the FWHM of the $^{13}$CO is high. However, as their observations do not only focus on the star-forming parts of Orion-B but on the whole region, the median $M$ decreases to $6$. In our case, observations were centred on two star-forming regions which explain the higher $M$. Such high-values of $M$ have been observed for instance toward Orion A \citep{gon19}, GMF38a \citep{wu18} or quiescent 70~$\mu$m clumps \citep{tra18a}. 

\subsection{Impact of the H{\,\sc{ii}} region on the layer}\label{sect:rdi}

\begin{figure}
  \centering
  \includegraphics[width=0.5\textwidth]{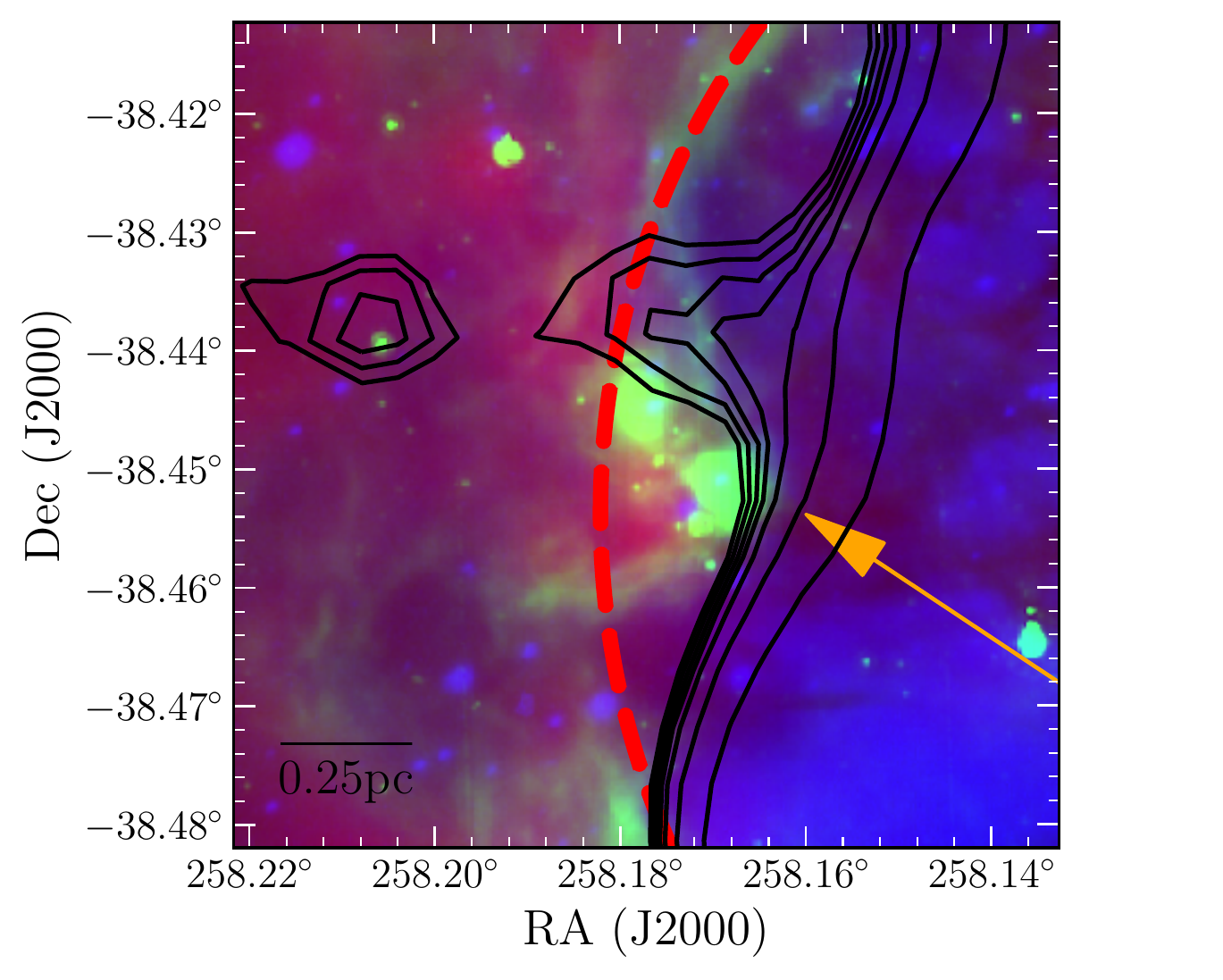}  
  \caption{Observation of region B from \textit{Spitzer} at 8~$\mu$m where the red dashed line represents the circle which follows the 8~$\mu$m emission, the black contours represent the free-free emission from SUMSS (contours from 0.01 to 0.04~Jy~beam$^{-1}$) and the orange arrow indicates the direction of the ionizing radiation. The clump with the tails are inside the circle.}
  \label{fig:Region_B_8um}
\end{figure}

Contrary to the rest of the ring, region B is the most intriguing due to its particular morphology. As seen on Fig.~\ref{fig:Region_B_8um}, the 8~$\mu$m observation shows a distorted emission as if a clump was pre-existing. This fact is particularly well-seen when following the distribution of 8~$\mu$m emission traced by the dashed red circle in Fig.~\ref{fig:Region_B_8um} where the clump and the bow structures are clearly seen inside the circle. The bow would be the result of lower density wings which would have been swept-up by the radiation compared to the central overdensity of the clump. The radio emission at 843~MHz from the Sydney University Molonglo Sky Survey (SUMSS), seen in black contours, also show a distortion around the clump of region B which indicates an interplay between the ionizing radiation and the clump. In the northern part of the clump, where the SUMSS emission is present, it can be also noted that an 8~$\mu$m arc is present, touching the last contour. The tails can also be observed on the velocity integrated image of C$^{18}$O (Fig.~\ref{fig:int_z}) but this is not really clear as the resolution of the CO observations ($\sim$18.2") is lower than \textit{Spitzer} 8~$\mu$m and \textit{Herschel} 70~$\mu$m observations.\\

Using radio continuum emission, the photon flux $\Phi$ impacting the clump of region B and the corresponding electron density $n_e$ can be computed with the following equation \citep{lef97,tho04}:

\begin{equation}\label{eq:photon_flux}
\Phi=1.24\times10^{10}\times \left[\frac{S_{\nu}}{\textrm{mJy}}\right]\times \left[\frac{T_e}{\textrm{K}}\right]^{0.35}\times \left[\frac{\nu}{\textrm{GHz}}\right]^{0.1}\times \left[\frac{\theta}{"}\right]^{-2}
\end{equation}

\begin{equation}\label{eq:ne}
n_e=122.21 \times \left(\frac{S_{\nu}T_e^{0.35}\nu^{0.1}}{\theta^{2}}\right)^{\frac{1}{2}} \times \left[\frac{\eta R_c}{\textrm{pc}}\right]^{\frac{-1}{2}}
\end{equation}

The electron temperature inside the {H\sc{ii}} region computed from \citet{bal11} gives $T_{\rm{e}}=8100~K$, in agreement with the range of $T_{\rm{e}}$ (7$\times 10^3$ to $10^4$~K) used in other works \citep{and15}. $\theta$ is the angular diameter where the flux is integrated, $R_c$ is the radius of clump impacted by the radiation and $\eta$ is the fraction of the clump which is photoionised. At 835~MHz, the flux is equal to 7~Jy in $\theta=506\arcsec$, $R_c$ is taken to be $\sim$0.18~pc and we assume the general value $\eta=0.2$ \citep{ber89}. The corresponding $\Phi$ at the interface between the H{\,\sc{ii}} region is equal to (8$\pm$1)$\times$10$^{9}$~cm$^{-2}$~s$^{-1}$. The value of the electron density $n_e$ at the edge of the clump is $\sim$(510$\pm$30)~cm$^{-3}$ which is far above the electron density value needed to form a Ionized Boundary Layer (IBL). Uncertainties have been estimated by using $T_e=7000-10000$~K and an uncertainty of 5\% for the flux \citep{mur07}. We have to note that the emission at 843~MHz can be optically thick and underestimated. Using the measurement from \citet{cas87} at 5~GHz for the whole region and equal to 8.3~Jy, $\Phi$ and $n_e$ increase to 1$\times$10$^{10}$~cm$^{-2}$~s$^{-1}$ and 600~cm$^{-3}$.\\

To understand if the pressure of the H{\,\sc{ii}} region could have pushed and compressed the clump of region B, we computed the pressure at the edge of the clump due to the ionization, $P_{ion}$, and the internal pressure of the clump, $P_{clump}$. The ionization pressure of an H{\,\sc{ii}} region is estimated with $P_{_{ion}}=2n_{e}T_{e}$ \citep{urq04}. The ionization pressure $P_{ion}/k_B$ is found to be (8$\pm$2)$\times$10$^{6}$~K~cm$^{-3}$. This value is similar to the one found toward the horshead nebula \citep{war06} with an O9.5 star at $\sim$3.5~pc of distance. The pressure of the clump is computed following P$_{clump}=\rho_{clump}\sigma ^{2}$ where $\rho_{clump}$ is the clump density and $\sigma$ ($\sim$1$\pm0.2$~km~s$^{-1}$) is the velocity dispersion of $^{13}$CO toward the clump. The pressure $P_{clump}/k_B$ is around (8$\pm$3)$\times$10$^{6}$~K~cm$^{-3}$.\\ 
If the clump was pre-existent without star-formation, the dispersion would mainly be thermal with $\sigma_{th}\sim 0.2$ (at 10~K), giving an upper limit of 1$\times$10$^{6}$~K~cm$^{-3}$ for $P_{clump}/k_B$. In this initial configuration, the clump of region B is firstly compressed by the H{\,\sc{ii}} region pressure, causing the gravitational collapse of the clump, the formation of stars and the increase of $P_{clump}$ through the increase of turbulence and temperature. When $P_{clump}\sim P_{ion}$, the highest density part of the clump stops to be compressed but the less dense part continues to be pushed, forming the wings and the bended shape. The ionization pressure might still be effective towards the low-density northern part of RCW~120, where a champaign flow is observed and toward the south center where bended structured can be observed on \textit{Herschel} observations.\\
Because $P_{clump}\sim P_{ion}$, \citet{tor15} concluded that the ionization pressure cannot create the cavity, making the C\&C scenario inconsistent. However, the expansion of the H{\,\sc{ii}} region might have been compressed the layer and stopped when $P_{clump}\sim P_{ion}$. This is consistent with the fact that we barely detect any motion of the ring, with an expansion velocity between 1.2 and 2.3~km~s$^{-1}$ \citep{and15}.\\

\section{Discussion}

\subsection{Dynamics of the region}

Both regions contains protostars, either at the beginning of their evolution (region A) or more advanced YSOs (region B), at different projected distance from the ionizing star and therefore, differently impacted by the UV radiation. The Mach number of region A and B are high compared to the literature, indicating that turbulence is significant in these regions.
In region A, this turbulence can be explained by the impact of UV radiation on the clump, by the ongoing star-formation and by the outflow toward the massive $\sim$300~$M_{\odot}$ \textit{Herschel} source. It can explain the mass of the fragments at 0.01~pc scale \citep{fig18}, in disagreement with the thermal Jeans mechanism.
In region B, the turbulence is less important and could be explained by the lowest impact of stellar feedback from low-mass stars. Indeed, the spectral profile is not broad and no tracers of shock have been detected using the MALT90 survey. In addition, the clump being farther away, the impact of UV radiation should be less significant.\\

We tried to understand the impact of UV radiation on the abundance of $^{13}$CO and C$^{18}$O, and on the photodissociation of C$^{18}$O. Using the mass derived from the \textit{Herschel} column density map, the abundance of $^{13}$CO and C$^{18}$O are found to be lower compared to the general ISM values (2.7$\times10^{-6}$ and 1.7$\times10^{-7}$, respectively, \citealt{gol97,mag88,pin08}). However, several bias have to be taken into account before drawing any conclusions about these values.\\
Firstly, the N($H_2$) mass differs from the one estimated by \citet{deh09} by a factor of 3 at most, hence the abundance would increase by the same factor. Secondly, $^{13}$CO being moderately thick toward the densest parts, the derived N($^{13}$CO) represents a lower limit. Accounting for this mass uncertainty on C$^{18}$O does not rule out a possible photodissociation.\\
The average ratio of $^{13}$CO to C$^{18}$O is equal to 4.6 and 7.4 for region A and B, respectively. These values are close to the value of 5.5 for the solar system and far from the high ratio observed towards Orion-A of 16.5 and the maximum of 33 \citet{shi14}. By looking at the ratio maps, we observed that the value is lower towards the densest part of the region and higher toward the edges. Unfortunately, as noted by \citet{shi14}, this ratio is affected by the beam filling factor which could be as low as 0.4 as well as being non uniform over the area due to the different structures observed \citep{par18}. The resolution of our observations does not allow us to conclude about the dissociation of C$^{18}$O towards the PDR of RCW~120.

\subsection{Induced star formation toward RCW~120}

\subsubsection{C\&C mechanism}

The interplay between the ionizing radiation from massive stars and the turbulent medium was analysed by \citet{tre12} using hydrodynamical simulations. They showed that the probability density function (PDF) of the gas can be used to trace the unperturbed and shocked gas. The PDF of the highest density clump (region A) is well-fitted by a power-law, showing the relation between the ionization pressure and the compression of this clump \citep{tre14}. Studies of \citet{tho12} and \citet{min13} also indicates that this clump is likely to be triggered by the H{\,\sc{ii}} region. This compression would have locally increase the density and lead to the gravitational collapse of the layer. However, it is not clear what is the mechanism responsible for the formation of this clump where most of all the massive cores are found.\\
As seen on Fig.\ref{fig:RCW120_APEX_mapping}, this clump does not seem to be pre-existent as the interface between the PDR and the H{\,\sc{ii}} region is not distorded. Therefore, the spatial distribution of the dust and gas in this region would be visually in agreement with the C\&C mechanism. 
We note that the CCC model would give the same ring-like dust distribution but star formation would not be the result of any compression as the ring would have formed before emission of UV radiation.\\

To better understand if the C\&C process could be at work, we compared the dynamical age of the {H\sc{ii}} region $t_{dyn}$, to the time needed for the layer to fragment $t_{frag}$. Such comparisons were already performed: towards S~233 \citep{lad15} and S~24 \citep{cap16}, the C\&C mechanism does not seem to be possible while toward Sh2-39 \citep{dur17}, Sh2-104 \citep{xu17}, Sh2-212 \citep{deh08}, Gum~31 \citep{dur15} or Sh~217 \citep{bra11}, the C\&C model appears to be plausible.\\
In this work, we estimated $t_{dyn}$ using the model of \citet{tre14}. We use the same set of equations as in their work, from \citet{marher05}:

\begin{equation}
N_{\rm{LyC}}=\left(\frac{7.603\times 10^{46}\rm{s}^{-1}}{b(\nu,T_e)}\right)
\left(\frac{S_{\nu}}{\rm{Jy}}\right)\left(\frac{T_{\rm{e}}}{10^4~\rm{K}}\right)^{-0.33}\left(\frac{D}{\rm{kpc}}\right)^2
\end{equation}
\begin{equation}
b(\nu,T_e)=1+0.3195{\rm{log}}\left(\frac{T_{\rm{e}}}{10^4~\rm{K}}\right)-0.2130\rm{log\left(\frac{\nu}{GHz}\right)}
\end{equation}

The thermal radio-continuum emission of RCW~120 is equal to 5.81 and 8.52~Jy at 8.35 and 14.35~GHz, respectively \citep{lan00}. The value of log(N$_{LyC}$) is found to be 48.14~s$^{-1}$. The radius of the bubble is taken to be 1.8~pc (277" at 1.34~kpc).

\begin{figure}
  \centering
  \includegraphics[width=.5\textwidth]{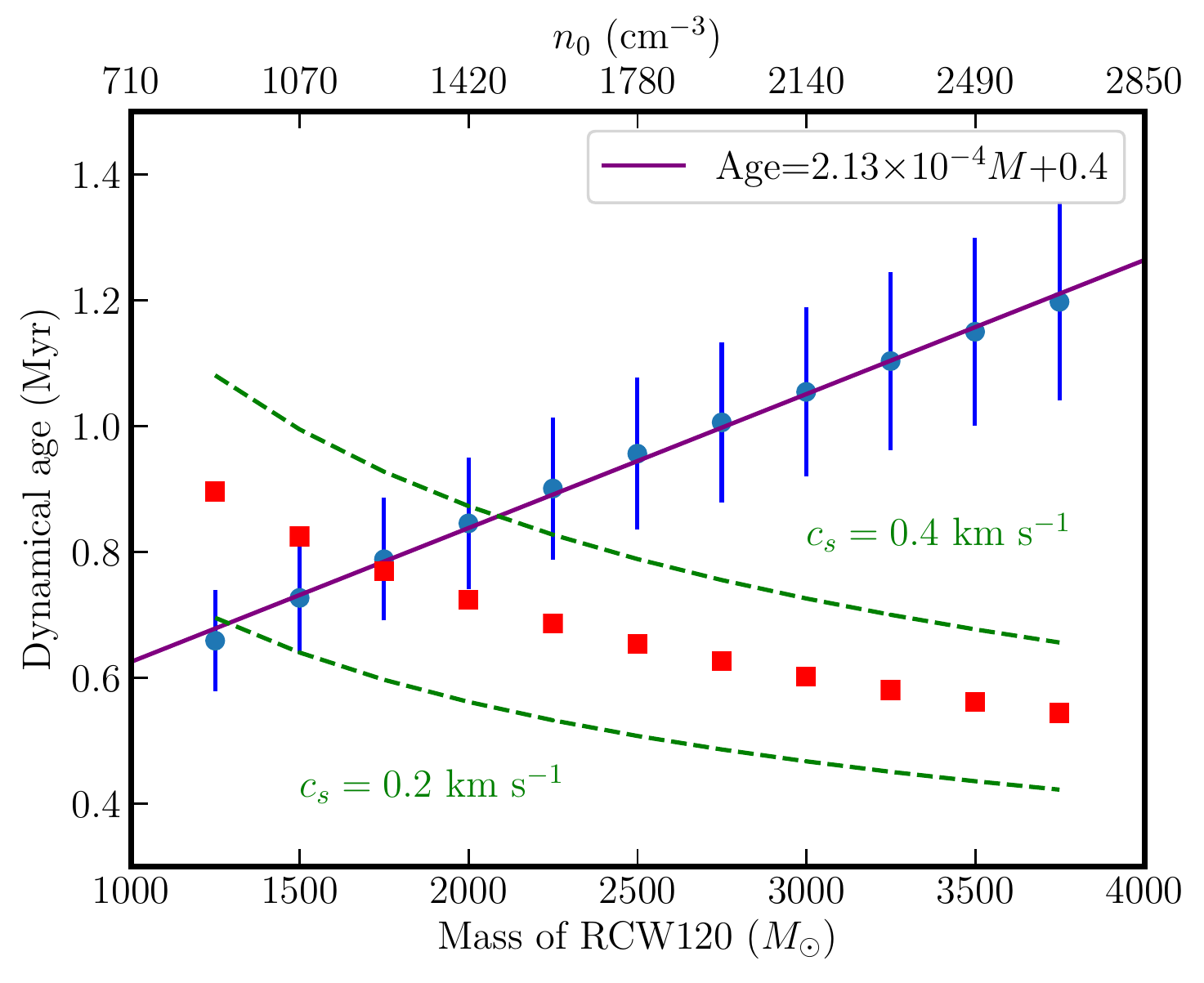}  
  \caption{Dynamical age as a function of the mass (bottom axis) - initial density (top axis) of RCW~120 (blue dot) with the uncertainty associated to each of them (blue bars) and the linear fit (purple line). Fragmentation timescale following the model of \citet{whi94b} (red squares) with the corresponding timescale assuming $c_s=0.2,0.4$~km~s$^{-1}$ (green dashed lines) and $c_s=0.3$, $N_{LyC}=10^{48}$~s$^{-1}$ (red squares)}
  \label{fig:dynamical_age}
\end{figure}

The last parameter needed is the initial density of the medium $n_0$. We estimated $n_0$ by assuming that all the mass was initially gathered in a sphere of $\sim$1.8~pc radius. The mass of RCW~120 at 870~$\mu$m, assuming $T=20$~K, is $\sim 2000$~$M_{\odot}$ \citep{deh09}. However, observations with APEX-LABOCA suffer from loss of large scale emission. Using the maps combined with \textit{Planck} data \citep{cse16} to correct from this emission loss , the mass of RCW~120 (contour of 0.6~Jy~beam$^{-1}$, $T=20$~K) increases to 2600~$M_{\odot}$. Using the column density map of RCW~120 \citep{fig17}, the mass of the layer is 6000~$M_{\odot}$ and increases to 10$^{4}$~$M_{\odot}$ if we consider the whole region.\\
The mass from LABOCA+\textit{Planck} (lower limit) and from N($H_2$) (upper limit) found give an initial density of 1.85$\times10^{3}$ and 7.13$\times10^{3}$~cm$^{-3}$, respectively. From the statistical study of \citet{pal17} with Hi-GAL \citep{mol10a}, H{\,\sc{ii}} regions in the Galactic plane have density ranging from $\sim$10 to $\sim$2300~cm$^{-3}$ and most of the simulations takes initial density between 1000 and 3000~cm$^{-3}$ \citep{art11,wal15,mac16,mar19}. Therefore, the initial density of 7.13$\times10^{3}$~cm$^{-3}$ seems too high compared to the usual values found toward H{\,\sc{ii}} regions.\\
In the model of \citet{tre14}, the input density is given by the average density at 1~pc. To compute the dynamical age, we used the nearest grid values to the estimations (8000~K, 1.8~pc and 10$^{48}$~s$^{-1}$). Uncertainties were estimated by using the grid values below and above the estimations. The results are plotted on Fig.~\ref{fig:dynamical_age}. The dynamical age as well as those taken from other works are listed in Tab.~\ref{tab:agercw120}. At the lowest density (1.9$\times 10^{3}$~cm$^{-3}$), the dynamical age of RCW~120 can be 2 to 4 times higher than the values usually found in the literature.

\begin{table}
\tiny
\caption{Age of RCW~120}
\label{tab:agercw120}      
\centering                          
\begin{tabular}{c|c|c}
\hline
\hline
Reference & $n_0$ & Age \\
 & ($\times 10^{3}$~cm$^{-3}$) & (Myr) \\
\hline
This work & $1.9-7.1$ & 0.96$\pm$0.25 \\
 &  & 0.75$\pm$0.13 \\
\citet{mar10}\tablefootmark{a} & & $<5$ \\
\citet{art11} & $1$ & $0.2$ \\
\citet{mar19} & $1-3$ & $0.23-0.42$ \\
\citet{zav07} & $3$ & $0.4$ \\
\citet{pav13} & $3-10$ & $0.17-0.32$ \\
\citet{aki17} & $1$ & $0.26-0.63$ \\
\hline
\hline
\end{tabular}
\tablefoot{
In \citet{mar10}, the estimation is obtained using isochrones
}
\end{table}

Previously, we mentioned that part of $N_{\rm{LyC}}$ could be absorbed by the dust, and this fraction can range from 25 to 50\% \citep{ino01}. Therefore, $N_{\rm{LyC}}=(1-f_0)N_{\rm{LyC}}^*$ where $f_0$ represents the absorption by dust is a lower-limit of the true $N_{\rm{LyC}}^*$. Following the calibration of O stars from \citet{mar05}, $N_{LyC}=10^{48.29}$~s$^{-1}$ for a O8V star. Compared to the previous estimation, up to 30\% of the ionizing photons are absorbed by the dust. Taking a flux of $10^{48.5}$ in the model, $t_{dyn}$ decrease to 0.75$\pm$0.13~Myr.

Having an estimation of$t_{dyn}$ allows a comparison with $t_{frag}$, the time needed for the layer to start fragmenting under the C\&C mechanism, using the analytical expression of \citet{whi94b}:

\begin{equation}\label{eq:whitworth}
t_{\rm{frag}}\sim1.56~\rm{Myr}~\left(\frac{c_s}{0.2~\rm{km}~\rm{s}^{-1}}\right)^{\frac{7}{11}}\left(\frac{N_{\rm{LyC}}}{10^{49}~s^{-1}}\right)^{\frac{-1}{11}}\left(\frac{n_0}{10^2~cm^{-3}}\right)^{\frac{-5}{11}}
\end{equation}

We have to note that the normalization constants used in \citet{whi94b} are extreme values, chosen to minimize the mass of the fragments. For the density normalization, a most realistic value proposed in the same work is 100~cm$^{-1}$, used in Eq.~\ref{eq:whitworth}, leading to a factor of 0.35 for $t_{frag}$. Such $n_0$ has been used by \citet{pal17} for instance.\\
On first approximation, $c_s$ depends on the temperature of the shell which is 28.3~K for region~A based on CO observations, giving $c_s=0.3$~km~s$^{-1}$. However, this is a lower limit as turbulence and sub-Lyman radiation leaking through the PDR could enhance it. $N_{LyC}$ does not have a lot of weight with respect to $t_{\rm{frag}}$, leading to a change of $\pm$0.3~Myr for a difference of one order of magnitude. An increase will lower $t_{\rm{frag}}$ as the column density threshold necessary for fragmentation would be reached faster. The dependence of $t_{\rm{frag}}$ with respect to the mass of RCW~120 (equivalently, $n_0$) is plotted on Fig.~\ref{fig:dynamical_age}. As for $N_{\rm{LyC}}$, a higher $n_0$ decreases $t_{frag}$ since the column density threshold is reached faster. We also computed $t_{frag}$ using $c_s=0.2-0.4$ to quantify the difference.\\
For a mass of 2600~$M_{\odot}$, $t_{frag}=0.5-0.7-0.8$~Myr (see Fig.~\ref{fig:dynamical_age}). Since $t_{dyn}<t_{\rm{frag}}$, the C\&C mechanism seems be the explanation for the fragmentation of the surrounding shell around the ionizing star. Simulations of \citet{aki17} and \citet{zav07} are also favorable to the C\&C mechanism at a density of 3$\times 10^{3}$~cm$^{-3}$. However, \citet{kir14} found the C\&C mechanism to be unlikely unless the density reaches $7\times 10^{3}$~cm$^{-3}$. This difference can explained by the choice of normalization for $n_0$. Indeed, the dynamical age in their work at $\sim 3000$~cm$^{-3}$ is 2~Myr, which would be reduced to 0.7~$Myr$ with a normalization of $100$~cm$^{-3}$. The comparison of $t_{dyn}$ and $t_{\rm{frag}}$ in this work would support the idea of triggering but, as we see in Tab.~\ref{tab:agercw120}, the age of RCW~120 in the literature might be as low as 0.17~Myr, which would make the C\&C inconsistent.\\
\citet{dal09} showed that, without pressure confinement, the thickness of the layer increases with time, as observed towards H{\,\sc{ii}} bubbles \citep{chu07}, but the simulations do not agree with the thin shell approximation used in the analytical model. Additionally, the magnetic field should also betaken into account \citep{fuk00}.\\

Therefore, conclusions from this model, even if it supports the idea of triggered star-formation, must be taken with caution. Detailed analysis of the interplay between the H{\,\sc{ii}} region and the PDR \citep{tre14,fig18,zav20} might be a better indicator but are also not exempt of some uncertainties \citep{dal15} regarding to induced star-formation.

\subsubsection{RDI process toward region B}

Several works have studied the impact of the photoionization pressure on pre-existing clumps, inducing the formation of stars. The study of \citet{urq09} on a sample of 45 clouds \citep{sug91,sug94} allow to understand the difference, in term of physical properties, between clouds where RDI is happening and where it is unlikely. For instance, an IBL through hydrogen recombination and a PDR through PAH emission at 8~$\mu$m should be observed. As UV radiation should heat the cloud, $T_{ex}^{12}$ should be higher and higher than $T_{ex}^{18}$. Moreover, some clouds triggered by the RDI mechanism show multiple components in CO, indicative of shocked/moving gas. In term of stars, as the ionization front propagates into the clump, sequential star formation might be observed. Additionally, RDI is thought to be efficient, leading in priority to an high-mass star or a cluster of intermediate mass stars \citep{sug91,mor08} toward the center of the clump \citep{kes03}. Estimations of the turbulence from NH$_3$ \citep{mor10} showed a clear bimodality with a higher turbulence in the triggered sample of BRCs.\\

On the H$\alpha$ observations from SHS (SuperCOSMOS H$\alpha$ Survey, \citealt{par05}), the H$\alpha$ emission stops where the 8~$\mu$m emission is located but no clear IBL is observed, despite a high $n_e$. On the other hand, the curved rim is very well seen at 8~$\mu$m, tracing the PDR and indicating a strong interaction between the clump and the UV radiation. It should be noted that, as Herbig Ae/Be stars may be present in region B (source 24 and 28), the PAH present around them can be partially be due their own FUV radiation \citep{seo17}.

As discussed previously (see Tab.~\ref{tab:T_LTE_RADEX}), $T_{ex}^{12}$ is found to be $\sim$21~K on average which is in agreement with \citet{urq09} for the sample of triggered BRCs. The values of $T_{ex}^{18}$ are found to be quite similar to $T_{ex}^{12}$ which might be explained by evolved YSOs inside region B \citep{deh09}, heating  the clump.

Several CO observations pointed out that multiple components, representing the dynamics, can be observed towards triggered BRCs. Different components ($-$12, $-$10 and $-1$~km~s$^{-1}$) are observed toward region B (Fig.~\ref{fig:spectra_fig_int_z}) but likely correspond to cloud emission on the line of sight (see Fig.~11 of \citealt{and15}). We have to note that multiple components observations is not a requirement for star formation to be triggered by RDI process. The sample of southern BRCs studied by \citet{urq09} was classified between spontaneous and triggered candidates based on the association with a PDR or an IBL. SFOs 51 and 59, part of the spontaneous sample show multiple components but SFOs 64 and 65 show only a single component while being considered as triggered candidates.

Based on the classification of YSOs in \citet{deh09}, sources in region B are Class I-II \citep{deh09} so star-formation is likely to be coeval in that clump. As the UV radiation interacts with the edge and propagates inside the clump, sequential star-formation or similar YSOs at the same location can be expected. From the analysis of the $L_{bol}-M_{env}$ diagram in \citet{fig17}, we also note that they are the most evolved YSOs in RCW~120 and were probably formed before the majority of the sources found in the PDR. This strengthens the idea that the clump in region B was pre-existing and star formation begins there when the rest of layer was still assembling.

The star formation efficiency is difficult to estimate since the stellar masses are not known. However, we can estimate the core formation efficiency (CFE) based on the mass of the cores and dust. Following \citet{fig17}, the total mass of the cores is 23~$M_{\odot}$ and the mass of the clump is between 90 and 200~$M_{\odot}$. The CFE varies between 12 and 26\% and since part of the envelope will be swept away during the formation of the stars, the final star formation efficiency will be lower than these values. Toward the Cepheus B population, \citet{get09} found that RDI is likely to have triggered star formation and that the SFE is between 35 and 55\%, well above the values found for region B. In the case of RCW~120, the H{\,\sc{ii}} region interaction might have formed stars in a shorter time but no high SFE is observed.\\

We compared our results with the RDI simulations developed by \citet{mia09} and, in particular, the cloud A of the first set ($n_0=2672$~cm$^{-3}$, $R=0.5$~pc). At the end of the simulation, the pressure of the clump is $\sim$10$^6$~K~cm$^{-3}$. A core is formed with a density 10$^{5}$~cm$^{-3}$, a temperature of 28~K, a mass of 15~$M_{\odot}$ and is found up to 0.3~pc from the surface layer. Compared to the region B, the pressure of the cloud is lower but may be due to the difference of initial density. Several cores have formed at the clump surface but we often observed multiple stars forming instead of one unique core at the top of the cloud if the BRC is not symmetrical \citep{sic14}. Regarding to the CFE, it rises to 38\% and is well above the value of 12 to 26\% estimated above.\\
Finally, the formation of the BRC into a A-type morphology and the collapse of the core takes about 0.4~Myr. This is in agreement since the dynamical age of RCW~120 is above this value. During the remaining 0.5~Myr after the formation of the BRC, the cores evolved into Class~I-II YSOs/stars and the ionization pressure increases the density of the clump, leading to a higher clump pressure. As already found by \citet{lef94}, \citet{mia09} showed that clouds located closer to the ionizing source will evolve to a type-A BRC. This is in agreement with the slightly curved clump of region B.\\
Compared to the simulations of \citet{kin15}, the time needed for the formation of a BRC is much lower ($\sim$0.1~Myr). However in that case, the total core mass produced is lower ($\sim$1.5$M_{\odot}$) and the resulting CFE is around 5\%.\\
Simulations of \cite{haw13} show a curved morphology of the CO distribution after the interaction with the H{\,\sc{ii}} region. This is observed at 8~$\mu$m but not in CO as the resolution might be too low ($\sim$18.2$"$). Depending on the strength of the ionizing flux and the viewing angle, the CO profiles can have multiple components, representing the dynamics of the clump. As we stressed before, CO profiles of triggered BRCs can show multiple components \citet{urq09, mor09} but this is not a requirement as it can depend on the viewing angle and ionizing flux strength.\\
The variation of the line profile symmetry parameter $\delta$ \citep{mar97} was also studied as a function of the viewing angle. Considering the profile of $^{12}$CO and C$^{18}$O, we found $\delta\sim0.2$, corresponding to an angle of $-$20$^{\circ}$  which indicate that the ionizing star should be in front of the dusty ring. This is in agreement with the PAH emission at 8~$\mu$m which appears face on while it would be unobservable if UV emission was coming from behind, as stated by \citet{urq09}.

\section{Conclusions}

In this work, we analysed the $^{12}$CO, $^{13}$CO and C$^{18}$O in the $J=3-2$ molecular transition toward two sub-regions located in the PDR of the Galactic H{\,\sc{ii}} region RCW~120. The region A corresponds to a high-mass clump where young high-mass cores were detected with {\textit{Herschel}} and ALMA while evolved low-mass YSOs were found in region B. Derivation of the velocity dispersion maps show an increase in both regions where star formation is observed and, assuming LTE, $T_{ex}$ is also found to increase toward the same locations. The estimated Mach number for both regions shows the supersonic motions inside the PDR, due to the impact of FUV and feedback from star-formation occurring there. The properties of the outflow toward the massive core of region A, traced by molecular transition from MALT90, are in good agreement with MSF regions from the literature.\\
We discussed the star formation with respect to the C\&C mechanism and found that the time needed for the layer to fragment is equivalent to the dynamical age of RCW~120. It appears also clear from other studies that the H{~\sc{ii}} region compressed the layer.\\
Toward region B, no IBL is observed compared to what is predicted by the high electron density value but PAH emission is observed Additionally, the radio emission engulfs the clump in region B and the 8~$\mu$m emission show wings and BRC A-type morphology. From simulations based on the RDI mechanism, the time needed for the ionization to form stars is in agreement with the dynamical age of the H{\,\sc{ii}} region. The higher pressure of the clump compared to the ionization pressure show that the compression of this clump stopped, in agreement with the low expansion velocity of the region. Therefore, region B appears as a good candidate for the RDI mechanism.

\begin{acknowledgements}
We thank the anonymous referee for useful comments and suggestions. LB and RF acknowledge support from CONICYT project Basal AFB-170002. MF has been supported by the National Centre for Nuclear Research (grant 212727/E-78/M/2018)
\end{acknowledgements}

%
   \bibliographystyle{aa} 
   \bibliography{../biblio} 
%

\clearpage
\begin{appendix} 

\begin{figure*}

\section{Pixel spectra}

  \centering    
  \includegraphics[width=1.\textwidth]{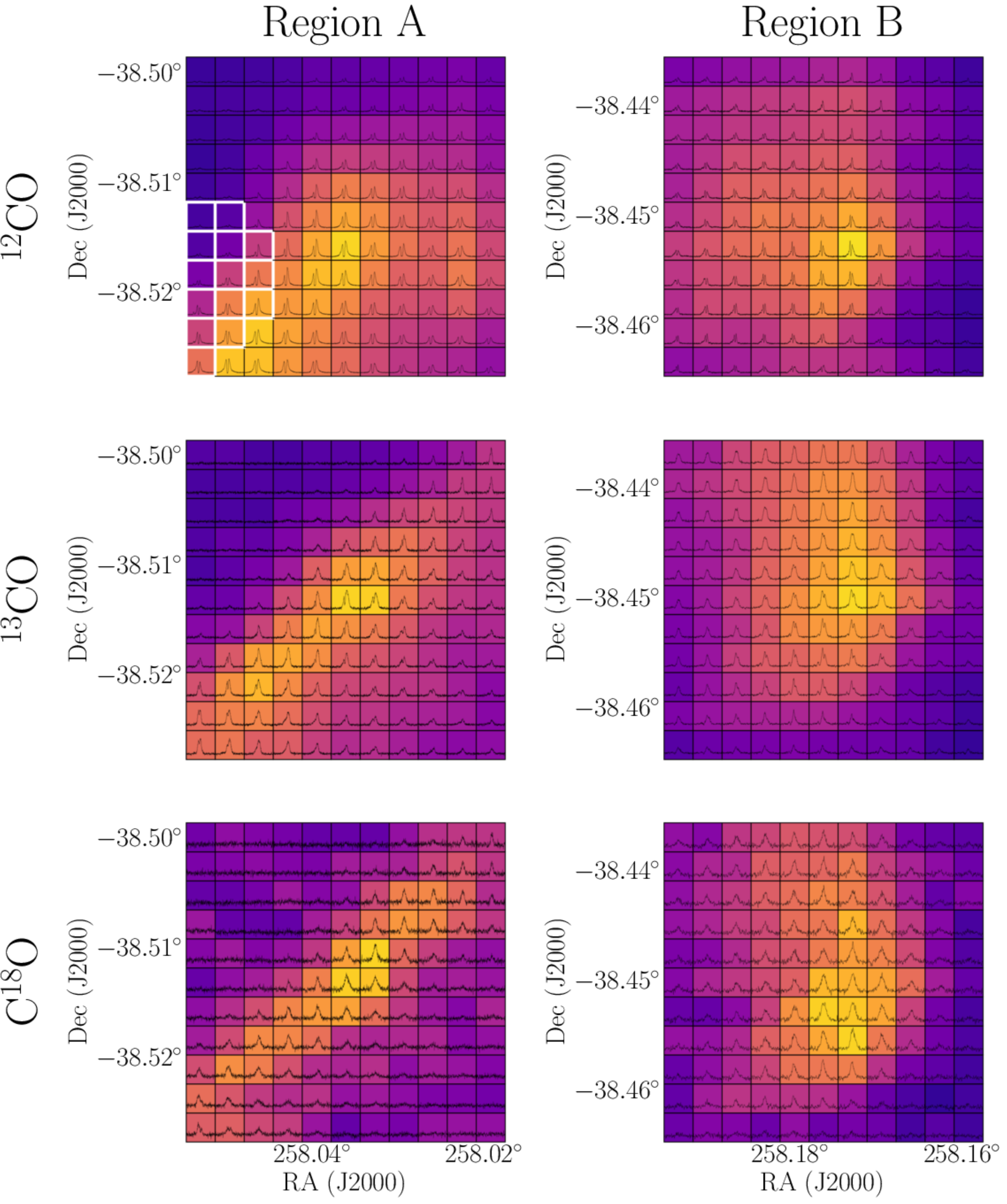}
  \caption{$^{12}$CO, $^{13}$CO and C$^{18}$O spectra at each pixel of the maps toward regions A (first column) and B (second column). Pixel spectra have the same axis range within the same map but differ from one region and isotopologue to another. Pixels with white edges indicate where a second component is seen at $-$15~km~s$^{-1}$.}
  \label{fig:spectra_fig_int_z}
\end{figure*}

\end{appendix}

\end{document}